%% file: main.tex
\def \arxivversion {0}

\if \arxivversion 0
    
    \documentclass[twoside,11pt]{article}
    \usepackage{blindtext}
    \usepackage{amsthm}
    \usepackage[preprint]{jmlr2e} 
    \usepackage{lastpage}
    \jmlrheading{27}{2026}{1-\pageref{LastPage}}{4/26}{4/26}{27-0000}{Peter A. Fisher and Anuradha M. Annaswamy}
    \ShortHeadings{Adapt and Stabilize, Then Learn and Optimize}{Fisher and Annaswamy}
    \firstpageno{1}
\else
    \documentclass{article}
    \usepackage[margin=1in]{geometry}
\fi

\input{utils}

    \hypersetup{hidelinks}

\title{Adapt and Stabilize, Then Learn and Optimize: A New Approach to Adaptive LQR}

\if \arxivversion 0

    \author{\name Peter A. Fisher \email pafisher@mit.edu \\
        \name Anuradha M. Annaswamy \email aanna@mit.edu \\
        \addr Department of Mechanical Engineering \\
        Massachusetts Institute of Technology \\
        Cambridge, MA 02139-4301, USA}
    \editor{My Editor}
\else
    \author{Peter A. Fisher and Anuradha M. Annaswamy}
    \date{}
\fi

\begin{document}

\maketitle

\input{abstract}

\if \arxivversion 0
\fi

\input{Sections/Introduction}
\input{Sections/Preliminaries}

\input{Sections/Adaptive_Control_Design}
\input{Sections/Algorithm}

\input{Sections/Analysis}

\input{Sections/Simulations}
\input{Sections/Related_Work}

\input{Sections/Future_Work}
\input{Sections/Conclusions}

\section*{Acknowledgements}

This work was supported by the Boeing Strategic University Initiative and by the Air Force Research Laboratory.



\newpage

\appendix

\input{Appendices/Random_Variables}
\input{Appendices/Comparator_System_Proofs}
\input{Appendices/Regret_Proof_new}
\input{Appendices/Additional_Simulations}


\vskip 0.2in
\bibliography{references}

\end{document}

%% file: utils.tex
\usepackage[utf8]{inputenc} 
\usepackage[T1]{fontenc}    
\usepackage{hyperref}       
\usepackage{url}            
\usepackage{booktabs}       
\usepackage{amsfonts}       
\usepackage{nicefrac}       
\usepackage{microtype}      
\usepackage{xcolor}         
\usepackage{amsmath}
\usepackage{amssymb}
\usepackage{mathtools}
\usepackage{graphicx}
\usepackage{float}
\usepackage{color}
\usepackage{transparent}
\usepackage{algorithm}
\usepackage{algorithmic}
\usepackage{subcaption}

\renewcommand{\vec}{\mathbf}

\renewcommand{\bar}{\overline}
\renewcommand{\hat}{\widehat}
\renewcommand{\tilde}{\widetilde}

\DeclareMathOperator{\argmin}{arg\,min}

\DeclareMathOperator*{\argminbelow}{arg\,min}

\newcommand{\Tr}{\mathrm{Tr}}
\newcommand{\rank}{\mathrm{rank}}

\newcommand{\ZMsubG}{\mathrm{ZMsubG}}

\newcommand{\dare}{\mathrm{dare}}
\newcommand{\dlqr}{\mathrm{dlqr}}
\newcommand{\bbR}{\mathbb{R}}
\newcommand{\bbN}{\mathbb{N}}
\newcommand{\bbZ}{\mathbb{Z}}
\newcommand{\bbE}{\mathbb{E}}
\newcommand{\bbP}{\mathbb{P}}

\newcommand{\bbC}{\mathbb{C}}

\newcommand{\calF}{\mathcal{F}}
\newcommand{\calO}{\mathcal{O}}
\newcommand{\calN}{\mathcal{N}}
\newcommand{\calE}{\mathcal{E}}

\newcommand{\calC}{\mathcal{C}}

    \renewcommand{\qed}{\hfill $\blacksquare$}
\newcommand{\proj}{\mathrm{proj}}

\newcommand{\Regret}{\mathrm{Regret}}
\newcommand{\indenti}[1]{\transparent{0}{#1}\transparent{1}}

\newcommand{\op}{\mathrm{op}}

    \newtheorem{theorem}{Theorem}

    \newtheorem{lemma}{Lemma}

    \newtheorem{proposition}{Proposition}

    \newtheorem{definition}{Definition}
    \theoremstyle{definition}
    \newtheorem{assumption}{Assumption}
    \newtheorem{remark}{Remark}

    \newtheorem{condition}{Condition}

\renewcommand{\algorithmiccomment}[1]{\bgroup\hfill\#~#1\egroup}

%% file: abstract.tex
\begin{abstract}
    This paper focuses on adaptive control of the discrete-time linear quadratic regulator (adaptive LQR). Recent literature has made significant contributions in proving non-asymptotic convergence rates, but existing approaches have a few drawbacks that pose barriers for practical implementation. These drawbacks include (i) a requirement of an initial stabilizing controller, (ii) a reliance on exploration for closed-loop stability, and/or (iii) computationally intensive algorithms. This paper proposes a new algorithm that overcomes these drawbacks for a particular class of discrete-time systems. This algorithm leverages direct model-reference adaptive control (direct MRAC) and combines it with an epoch-based approach in order to address the drawbacks (i)-(iii) with a provable high-probability regret bound comparable to existing literature. Simulations demonstrate that the proposed approach yields regrets that are comparable to those from existing methods when the conditions (i) and (ii) are met, and yields regrets that are significantly smaller when either of these two conditions is not met.
\end{abstract}

%% file: Sections/Introduction.tex
\section{Introduction} \label{sec:introduction}


Over the past decade and a half, a large body of work has developed on characterizing finite-time convergence guarantees for the adaptive LQR problem.
One can in fact trace the study of adaptive LQR back to Astr\"om's study of self-tuning regulators \citep{astrom1973selftuning} in the 1970s. Since then, the problem has had an extensive history. 
The more recent body of work in \citep{abbasi2011,ibrahimi2012sparseadaptiveLQR,Dean_2018,cohen2019SDPdare,mania2019CEefficient,simchowitz2020naive,lale2022reinforcement,sarker2023accurate} has generally focused mainly on characterizing theoretically achievable sample complexities and regret bounds: see Section \ref{sec:related_work} for a literature review. However, as suggested by \cite{lale2022reinforcement}, there is interest in the adaptive LQR community in moving towards algorithms that could lead to practical adaptive control on a physical system. There are clearly several hurdles to be overcome on the path to practicality. In particular, this paper addresses three drawbacks which make existing approaches unsuitable for practical implementation:
\renewcommand{\labelenumi}{(\roman{enumi})}
\begin{enumerate}
    \item a requirement of an initial stabilizing controller, which implies either open-loop stability or very good prior knowledge of the true parameters;
    \item a reliance on exploration, which may be limited by safety or actuator constraints, for closed-loop stability; and/or
    \item periodic solution of a non-convex constrained optimization problem, which may not be feasible in real time.
\end{enumerate}
The remainder of this paper proposes and analyzes a new adaptive LQR algorithm that addresses drawbacks (i)-(iii). The key to this approach is direct adaptive control, which is computationally efficient and achieves closed-loop stability with no requirements on an initial stabilizing controller or exogenous excitation.

\subsection{Problem statement}

We consider adaptive control of a discrete linear time-invariant plant given by
\begin{equation} \label{eqn:plant}
    x_{t+1} = A_*x_t + B_*u_t + w_{t+1}
\end{equation}
where $x_t \in \bbR^n$ is the (fully measurable) state, $u_t \in \bbR^m$ is the input, and $w_t \in \bbR^n$ is i.i.d noise with $w_t \sim \ZMsubG(\sigma_w^2I_n | \calF_t)$ (see Definition \ref{def:ZMsubG_vector} and Remark \ref{rmk:conditional_subG}) and $\bbE[w_{t+1}w_{t+1}^\top | \calF_t] = \Sigma_w$, where $\calF_t$ is a filtration on $x_t, u_t, w_t$. The initial condition $x_0$ is assumed deterministic. The dynamics $(A_*, B_*)$ is controllable but unknown, subject to the following assumptions:
\begin{assumption}[Matched Uncertainties] \label{asn:matching_condition}
    For a known Schur-stable dynamics pair $(A_m, B_m)$ where $B_m$ has full column rank, there exist matrices $\Theta_{A*} \in \bbR^{m \times n}$, $\Theta_{B*} \in \bbR^{m \times m}$ such that
    \begin{equation} \label{eqn:matching_condition}
        A_m = A_* + B_m\Theta_{A*}, \quad B_* = B_m\Theta_{B*}.
    \end{equation}
\end{assumption}
\begin{remark}
    Matched uncertainties are a common assumption in the direct adaptive control literature \citep{Narendra05,Goodwin_1984}. Many physical systems satisfy this assumption in continuous time. In discrete time, any dynamical system expressed as an ARMA model or a state-space model in controllable canonical form automatically satisfies this assumption. Additionally, if a continuous-time system $(A_*, B_*)$ satisfies \eqref{eqn:matching_condition} for a known Hurwitz pair $(A_m, B_m)$, then it can be shown that the overall uncertainty in the dynamics when discretized with a small time step $\Delta t$ is dominated by matched uncertainties, in the sense that the resulting discrete-time matched uncertainties have magnitude $\calO(\Delta t)$ and any unmatched uncertainties have magnitude $\calO(\Delta t^2)$.
\end{remark}
\begin{assumption}[Bounds on the Unknown Parameters] \label{asn:params_bounded}
    For the matrices $\Theta_{A*}$ and $\Theta_{B*}$ satisfying \eqref{eqn:matching_condition}, there exist known convex, compact sets $S_A \subset \bbR^{m \times n}$ and $S_B \subset \bbR^{m \times m}$ respectively, such that:
    \begin{enumerate}
        \item $\Theta_{A*} \in \mathrm{int}(S_A)$ and $\|\Theta\|_\op \leq a_{max}$ for all $\Theta \in S_A$; and
        \item $\Theta_{B*} \in \mathrm{int}(S_B)$, and $\det(\Theta) \neq 0$, $\|\Theta\|_\op \leq b_{max}$, and $\|\Theta^{-1}\|_\op \leq \frac{1}{b_{min}}$ for all $\Theta \in S_B$
    \end{enumerate}
    for some constants $a_{max}, b_{max}, b_{min} > 0$, where $\mathrm{int}(S)$ denotes the interior of $S$.
\end{assumption}
\begin{remark}
    $S_A$ and $S_B$ are needed in the algorithm as the parameter estimates $\hat{\Theta}_{At}$ and $\hat{\Theta}_{Bt}$ will be projected to them at every time step. This assumption is written to be as general as possible, but there are clear special cases that satisfy it. Assumption \ref{asn:params_bounded}.(i) simply asks for a known upper bound on $\|\Theta_{A*}\|_2$. Assumption \ref{asn:params_bounded}.(ii) asks for a known upper bound on $\|\Theta_{B*}\|_2$ and some additional a priori knowledge of $\Theta_{B*}$, including that it is invertible.
    
    A common special case is the setting where $\Theta_{B*}$ is diagonal with nonzero diagonal elements. In this case, compactness and convexity of $S_B$ requires knowledge of the signs of all diagonal elements - which is a typical assumption in direct adaptive control \citep{Narendra05} - and requires the magnitudes of all diagonal elements to lie in the range $[b_{min}, b_{max}]$ for known $b_{min}, b_{max} > 0$. Then, $S_B$ is the set of all diagonal $m \times m$ matrices whose diagonal elements are upper- and lower-bounded appropriately. This set is compact and convex, and thus one can construct a projection operator.

\end{remark}

Assumptions \ref{asn:matching_condition}-\ref{asn:params_bounded} will be assumed for all theoretical results in this work. The goal of LQR is to minimize the infinite-time control cost
\begin{equation} \label{eqn:lqr_optimal_cost}
    J_* = \min_{\{u_t\}_{t=0}^\infty} \lim_{T \to \infty} \bbE\left[\frac{1}{T}\sum_{t = 0}^{T-1} x_t^\top Qx_t + u_t^\top Ru_t\right]\ \mathrm{s.t.\ dynamics\ in\ \eqref{eqn:plant}}
\end{equation}
given cost matrices $Q = Q^\top \geq 0$ and $R = R^\top > 0$. It is well-known
\citep[see e.g.,][]{stengel1994optimal}
that the optimal controller is given by $u_t = K_*x_t$, where
\begin{gather}
    K_* = -(R + B_*^\top P_*B_*)^{-1}B_*^\top P_*A_*, \label{eqn:opt_K} \\
    P_* = A_*^\top P_*A_* - A_*^\top P_*B_*(R + B_*^\top P_*B_*)^{-1}B_*^\top P_*A_* + Q. \label{eqn:opt_dare}
\end{gather}
Equation \eqref{eqn:opt_dare} is known as the Discrete Algebraic Riccati Equation (DARE). We will subsequently denote $K_*$ calculated as in \eqref{eqn:opt_K}-\eqref{eqn:opt_dare} as $K_* = \dlqr(A_*, B_*, Q, R)$. In this work, as $A_*$ and $B_*$ are unknown, we seek to minimize the regret given by
\begin{equation} \label{eqn:regret_definition}
    \Regret(T) = \sum_{t = 0}^{T-1} \left(x_t^\top Qx_t + u_t^\top Ru_t - J_*\right).
\end{equation}
Finally, in this work, as in \citep{simchowitz2020naive}, we will require $Q \succeq I_n$, $R \succeq I_m$.

\subsection{Organization of the paper}

Section \ref{sec:preliminaries} lays out some preliminary results from the existing literature on sub-Gaussian random variables, quantifying excitation using spectral lines, and system identification via weighted recursive least squares. Section \ref{sec:algorithm} introduces our algorithm, MRAC-LQR, and Section \ref{sec:analysis} provides the theoretical analysis, including our main results on stability and regret. Section \ref{sec:simulations} compares the performance of MRAC-LQR to existing methods in simulation. Finally, Section \ref{sec:related_work} gives an overview of the existing literature on adaptive LQR, and we conclude with suggestions for several extensions of this work.

\subsection{Notation}

Given a dynamics pair $(A, B)$ and cost matrices $Q$ and $R$, the solution to the DARE is denoted as $\dare(A, B, Q, R)$ and the optimal feedback gain is denoted as $\dlqr(A, B, Q, R)$. Additionally, for a vector $a \in \bbR^d$, $\mathrm{diag}(a) \in \bbR^{d \times d}$ is the diagonal matrix with the elements of $a$ on the diagonal. We denote the operator and Frobenius norms of a matrix $A$ as $\|A\|_\op$ and $\|A\|_F$ respectively, and we denote the $d$-dimensional identity matrix as $I_d$.

%% file: Sections/Preliminaries.tex
\section{Preliminaries} \label{sec:preliminaries}

In this section, we provide some useful preliminary results from other work related to random variables and parameter convergence of recursive least squares. Additional useful preliminaries may be found in Appendix \ref{app:preliminaries}.


\subsection{Sub-Gaussian random variables and spectral lines}

In this work, we consider sub-Gaussian noise and employ a deterministic sum of sinusoids as an exploratory signal for parameter learning. We first require the following definitions:

\begin{definition}[\citealp{Vershynin2019}] \label{def:subG_variable}
    A random variable $X$ is sub-Gaussian with norm $\|X\|_{\psi_2}$ if the quantity
    \begin{equation} \label{eqn:subG_norm}
        \|X\|_{\psi_2} = \inf\{c > 0 : \bbE[\exp(X^2/c^2)] \leq 2\}
    \end{equation}
    exists and is finite.
\end{definition}

\begin{definition}[\citealp{Vershynin2019}] \label{def:ZMsubG_variable}
    A zero-mean random variable $X$ is sub-Gaussian with variance proxy $\sigma^2$ and is denoted $X \sim \ZMsubG(\sigma^2)$ if
    \begin{equation} \label{eqn:ZMsubG_variable}
        \bbE[\exp(\lambda X)] \leq \exp(\lambda^2\sigma^2/2)\ \forall \lambda > 0.
    \end{equation}
\end{definition}

\begin{remark}
    Definition \ref{def:ZMsubG_variable} is a special case of Definition \ref{def:subG_variable} -- see Lemma \ref{lem:ZMsubG_is_subG}.
\end{remark}

\begin{remark} \label{rmk:conditional_subG}
    If a random variable $X$ satisfies \eqref{eqn:ZMsubG_variable} with the expectation conditioned on a filtration $\calF_t$, we denote this as $X \sim \ZMsubG(\sigma^2 | \calF_t)$.
\end{remark}

\begin{definition}[\citealp{Pisier2016}] \label{def:ZMsubG_vector}
    A zero-mean complex-valued random vector $X \in \bbC^d$ is sub-Gaussian with variance proxy $\sigma^2$ and is denoted $X \sim \ZMsubG(\sigma^2I_d)$ if, for any unit vector $z \in \bbR^d$, $\|z\| = 1$, we have $\Re(z^\top X) \sim \ZMsubG(\sigma^2)$ and $\Im(z^\top X) \sim \ZMsubG(\sigma^2)$.
\end{definition}

\begin{definition}[\citealp{Vershynin2019}] \label{def:subE_variable}
    A random variable $X$ is sub-exponential with norm $\|X\|_{\psi_1}$ if the quantity
    \begin{equation} \label{eqn:subE_norm}
        \|X\|_{\psi_1} = \inf\{c > 0 : \bbE[\exp(|X|/c)] \leq 2\}
    \end{equation}
    exists and is finite.
\end{definition}

A few useful properties of sub-Gaussian
random variables and vectors are detailed in Appendix \ref{app:random_variables}.
The definitions above are useful in defining the concepts of a sub-Gaussian spectral line and an expected information matrix:
\begin{definition}[\citealp{sarker2023accurate}] \label{def:subG_spectral_line}
    A sequence $\{\phi_t\}_{t \geq 0} \in \bbR^d$ is said to have a sub-Gaussian spectral line from time $t_0$ to time $t_0+T_0-1$ with frequency $\omega_0$, amplitude $\bar{\phi}(\omega_0) \in \bbC^d$, and radius $\sigma$ if
    \begin{equation} \label{eqn:subG_spectral_line}
        \frac{1}{T_0}\sum_{t=t_0}^{t_0+T_0-1} \phi_te^{-i2\pi\omega_0t} - \bar{\phi}(\omega_0) \sim \ZMsubG\left(\frac{\sigma^2}{T_0}I_d\right).
    \end{equation}
\end{definition}
\begin{definition}[\citealp{sarker2023accurate}] \label{def:expected_info_matrix}
    Suppose that the sequence $\{\phi_t\}_{t \geq 0} \in \bbR^d$ has $d$ sub-Gaussian spectral lines from time $t_0$ to time $t_0+T_0-1$ with frequencies $\omega_1, \dots, \omega_d$ and amplitudes $\bar{\phi}(\omega_1), \dots, \bar{\phi}(\omega_d)$. Then, the expected information matrix is defined as
    \begin{equation} \label{eqn:expected-info_matrix}
        \bar{\Phi} = [\bar{\phi}(\omega_1), \cdots, \bar{\phi}(\omega_d)] \in \bbC^{d \times d}.
    \end{equation}
\end{definition}

Finally, this work will make use of two key results relating a deterministic sinusoidal input to excitation in a stable LTI system perturbed by sub-Gaussian noise.
\begin{proposition}[Adapted from \citealp{sarker2023accurate}] \label{prop:spectral_line_from_input}
    Consider a stable discrete-time LTI system given by $x_{t+1} = Ax_t + Bu_t + w_{t+1}$, $x_t, w_t \in \bbR^n$, $u_t \in \bbR^m$, with arbitrary initial conditions and $w_t \sim \ZMsubG(\sigma_w^2I_n)$ i.i.d. Suppose that the input is chosen as $u_t = Kx_t + r_t$ such that $A_K := A + BK$ is Schur-stable. Define $\phi_t := [-x_t^\top, u_t^\top]^\top$. Then, if $\{r_t\}_{t \geq 0}$ has a sub-Gaussian spectral line from time $t_0$ to time $t_0+T_0-1$ with frequency $\omega_0$, amplitude $\bar{r}(\omega_0)$, and radius $0$, then $\{\phi_t\}_{t \geq 0}$ has a sub-Gaussian spectral line from time $t_0$ to time $t_0+T_0-1$ with frequency $\omega_0$, amplitude $\bar{\phi}(\omega_0) = \begin{bmatrix} (e^{j\omega_0}I_n - A_K)^{-1}B \\ K(e^{j\omega_0}I_n - A_K)^{-1}B + I_m \end{bmatrix}\bar{r}(\omega_0)$, and radius $\sigma_\phi = \max\{1, \|K\|_\op\}\|(e^{j\omega_0}I_n - A_K)^{-1}\|_\op\sigma_w$.
\end{proposition}
\begin{proposition}[Adapted from \citealp{sarker2023accurate}] \label{prop:FE_from_spectral_lines}
    Suppose that the sequence $\{\phi_t\}_{t \geq 0} \in \bbR^d$ has $d$ sub-Gaussian spectral lines from time $t_0$ to time $t_0+T_0-1$ with frequencies $\omega_1, \dots, \omega_d$, linearly independent amplitudes $\bar{\phi}(\omega_1), \dots, \bar{\phi}(\omega_d)$, and maximum radius $\sigma$. Choose any $\alpha \in (0, \|\bar{\Phi}^{-1}\|_\op^{-2})$ and $\delta \in (0, 1)$. Then, with probability at least $1 - \delta$, if $T_0 \geq \frac{d\sigma^2\ln(9^{2d}/\delta)}{c(\|\bar{\Phi}^{-1}\|_\op^{-2} - \alpha)}$ for a universal constant $c > 0$,
    \begin{equation} \label{eqn:FE_from_spectral_lines}
        \frac{1}{T_0}\sum_{t = t_0}^{t_0+T_0-1} \phi_t\phi_t^\top \geq \frac{\alpha}{d}I.
    \end{equation}
\end{proposition}

\subsection{Recursive least squares and confidence sets}

Consider the online linear regression problem
\begin{equation} \label{eqn:online_regression}
    y_{t+1} = \Theta_*\phi_t + \eta_{t+1},
\end{equation}
where $\phi_t \in \bbR^d$ and $y_t \in \bbR^m$ are measured, $\eta_t \sim \ZMsubG(\sigma_\eta^2I_m | \calF_t)$ is unmeasured noise, where $\calF_t$ is a filtration on $\phi_t, y_t, \eta_t$, and $\Theta_* \in \bbR^{m \times d}$ is unknown. Suppose that, given $\{\phi_s\}_{0 \leq s \leq t}$ and $\{y_s\}_{1 \leq s \leq t+1}$, an estimate $\Xi_{t+1}$ of $\Theta_*$ is constructed using the following recursive least squares (RLS) adaptive law (see e.g., \citealp{Goodwin_1984}):
\begin{subequations}
    \begin{align}
        \Sigma_{t+1}^{-1} &= \Sigma_t^{-1} + \phi_t\phi_t^\top \label{eqn:RLS_cov_inv_update} \\
        \Sigma_{t+1} &= \Sigma_t - \frac{\Sigma_t\phi_t\phi_t^\top\Sigma_t}{1 + \phi_t^\top\Sigma_t\phi_t} \label{eqn:RLS_covariance_update} \\
        \Xi_{t+1} &= \Xi_t + (y_{t+1} - \Xi_t\phi_t)\phi_t^\top\Sigma_{t+1} \label{eqn:RLS_param_update}
    \end{align}
\end{subequations}
where \eqref{eqn:RLS_cov_inv_update} and \eqref{eqn:RLS_covariance_update} are equivalent by Sherman-Morrison, and $\Sigma_0^{-1} = \lambda I_d > 0$ for simplicity. It can be easily shown that $\Xi_t$ satisfies
\begin{equation}
    \Xi_t = \argminbelow_{\Xi \in \bbR^{m \times d}}\left[\lambda\|\Xi  - \Xi_0\|_F^2 + \sum_{s=0}^{t-1} \|y_{s+1} - \Xi\phi_s\|^2\right]
\end{equation}
Furthermore, let $\Theta_* \in S_\Theta$ for a known compact set $S_\Theta$. Then, \cite{abbasi2011} show how to construct a sequences of confidence sets which contain $\Theta_*$ with high probability as follows. Fix any $\delta \in (0, 1)$, and define
\begin{align}
    \beta_\delta(\Sigma_t) &= \left(m\sigma_\eta\sqrt{2\ln\left(\frac{\det(\Sigma_t^{-1})^{1/2}}{\delta\lambda^{d/2}}\right)} + \sqrt{\lambda}\max_{\Theta \in S_\Theta}\|\Xi_0 - \Theta\|_F\right)^2, \label{eqn:beta_delta} \\
    \calC_\delta(\Xi_t, \Sigma_t) &= \left\{\Theta \in \bbR^{m \times d} : \Tr[(\Theta - \Xi_t)\Sigma_t^{-1}(\Theta - \Xi_t)^\top] \leq \beta_\delta(\Sigma_t)\right\}. \label{eqn:C_delta}
\end{align}
The following result from \citep{abbasi2011} will play a key role in the optimality result in Theorem \ref{thm:regret}:
\begin{proposition}[\citealp{abbasi2011}] \label{prop:confidence_sets}
    Denote the event on which $\Theta_* \in \calC_\delta(\Xi_t, \Sigma_t)\ \forall t \geq 0$ as $\calE_1$. Then, $\calE_1$ holds with probability at least $1 - \delta$.
\end{proposition}

Additionally, consider any sequence of matrices $\{\Xi_t'\}_{t \geq 0}$ satisfying $\Xi_t' \in S_\Theta \cap \calC_\delta(\Xi_t, \Sigma_t)\ \forall t \geq 0$. The following result, proven in Appendix \ref{app:confidence_set_convergence}, is a crucial property of the confidence sets in \eqref{eqn:beta_delta}-\eqref{eqn:C_delta}:
\begin{proposition} \label{prop:confidence_set_convergence}
    Suppose that $\|\phi_t\|$ is a sub-Gaussian random variable with $\|\|\phi_t\|\|_{\psi_2}$ uniformly bounded in time. Then, on the event $\calE_1$, $\{\Xi_t'\}_{t \geq 0}$ satisfies $\sum_{t=0}^{T-1} \|(\Xi_t' - \Theta_*)\phi_t\|^2 \leq \tilde{O}(\sqrt{T})$ with probability one.
\end{proposition}

%% file: Sections/Adaptive_Control_Design.tex
\section{Adaptive control with time-varying reference dynamics} \label{sec:adaptive_control}

In this section, we introduce an adaptive control design for model-reference adaptive control with time-varying reference dynamics. In particular, we consider a reference model which varies in epochs: the dynamics are constant over the course of an epoch, and experience a jump from one epoch to the next. Let $t$ denote the time step, $k$ denote the epoch number, $t_k$ denote the first time step in epoch $k$ with $t_0 = 0$, $k_t$ denote the number of the epoch containing time step $t$, and $T_k = t_{k+1} - t_k$ denote the number of time steps contained in epoch $k$. The time-varying comparator system is given by
\begin{equation} \label{eqn:comparator_system}
    x_{c(t+1)} = A_{mk_t}x_{ct} + B_mr_t + w_{t+1},
\end{equation}
where $x_{c0} = x_0$, $\{w_t\}_{t \geq 1}$ is the same noise realization that perturbs the plant in \eqref{eqn:plant}, $A_{m0} = A_m$ and $(A_m, B_m)$ is the dynamics pair in Assumption \ref{asn:matching_condition}, and $\{r_t\}_{t \geq 0}$ is a bounded reference input satisfying $\|r_t\| \leq R < \infty\ \forall t \geq 0$.
\begin{remark}
    Traditionally, the term \textit{reference model} refers to a dynamical system which represents the desired closed-loop dynamics and whose dynamics and state are available for use in the control design. As $\{w_t\}_{t \geq 1}$ is unobserved, $x_{ct}$ is not available for use in the controller, and thus we refer to \eqref{eqn:comparator_system} as a \textit{comparator system} rather than a reference model. We introduce the notion of a comparator system as it is useful in our regret analysis later on.
\end{remark}
Algorithm \ref{alg:mrac-lqr} will show how to choose $A_{mk}$ for each $k \geq 1$ such that the adaptive controller achieves the optimality result in Theorem \ref{thm:regret}. For now, we will suppose that $\{A_{mk}\}_{k \geq 0}$ and $\{T_k\}_{k \geq 0}$ satisfy the following conditions:
\begin{condition} \label{cond:TVRM}
    For all $k \geq 0$, let:
    \begin{enumerate}
        \item $A_{mk} = A_m + B_m\Delta_{mk}$ for some $\Delta_{mk} \in \bbR^{m \times n}$ with $(A_m, B_m)$ as in Assumption \ref{asn:matching_condition},
        \item $\exists \Delta_{max} < \infty$ such that $\|\Delta_{mk}\|_\op \leq \Delta_{max}$,
        \item $\exists \gamma_m > 1$, $\lambda_m \in (0, 1)$ such that $\|A_{mk}^i\|_\op \leq \gamma_m\lambda_m^i\ \forall i \geq 0$,
        \item $\exists C_T > 0$ such that $T_k \geq C_T(k+1)$.
    \end{enumerate}
\end{condition}
\noindent Note that $\Delta_{m0} = 0$. The following result, proven in Appendix \ref{app:TVRM_stability}, guarantees the time-varying comparator system's stability:
\begin{theorem} \label{thm:TVRM_stability}
    Let Condition \ref{cond:TVRM} hold. Then, for the time-varying comparator system in \eqref{eqn:comparator_system}, $\exists \gamma_1 > 1, \lambda_1 \in (0, 1)$ such that $\|x_{ct}\|$ is a sub-Gaussian random variable with
    \begin{equation} \label{eqn:x_ct_subG_norm}
        \|\|x_{ct}\|\|_{\psi_2} \leq \frac{\gamma_1\lambda_1^t\|x_0\|}{\sqrt{\ln(2)}} + \gamma_1\frac{1 - \lambda_1^t}{1 - \lambda_1}\left(\frac{\|B_m\|_\op R}{\sqrt{\ln(2)}} + C_2'\sqrt{n}\sigma_w\right)
    \end{equation}
    for all $t \geq 0$, where $C_2' > 0$ is the universal constant from Lemma \ref{lem:ZMsubG_is_subG}.
\end{theorem}

\subsection{Adaptive control design} \label{subsec:adaptive_control_design}

We propose the following controller:
\begin{equation} \label{eqn:control_input}
    u_t = \hat{\Theta}_{Bt}^{-1}((\hat{\Theta}_{At} + \Delta_{mk_t})x_t + r_t),
\end{equation}
where $\hat{\Theta}_{At}$ and $\hat{\Theta}_{Bt}$ are estimates of $\Theta_{A*}$ and $\Theta_{B*}$ respectively from Assumption \ref{asn:matching_condition}. To derive a parameter estimation law, first note that, with some slight manipulations using Assumption \ref{asn:matching_condition}, one can easily rewrite the plant \eqref{eqn:plant} in the form \eqref{eqn:online_regression} with
\begin{equation} \label{eqn:online_regression_definitions}
    \begin{gathered}
        y_{t+1} = (B_m^\top B_m)^{-1}B_m^\top(x_{t+1} - A_mx_t), \quad \eta_{t+1} = (B_m^\top B_m)^{-1}B_m^\top w_{t+1}, \\
        \Theta_* = [\Theta_{A*}, \Theta_{B*}], \quad \phi_t = [-x_t^\top, u_t^\top]^\top.
    \end{gathered}
\end{equation}
It is easy to verify that $\eta_t \sim \ZMsubG(\sigma_\eta^2I_m | \calF_t)$ with $\sigma_\eta = \|(B_m^\top B_m)^{-1}\|_\op\sigma_w$.
Define $\hat{\Theta}_t = [\hat{\Theta}_{At}, \hat{\Theta}_{Bt}]$. Then, we produce $\hat{\Theta}_{At}$ and $\hat{\Theta}_{Bt}$ from the following adaptive law:
\begin{equation} \label{eqn:adaptive_law}
    \hat{\Theta}_{t+1} = \proj_{Z_{t+1}}(\hat{\Theta}_{t+1}'), \quad \hat{\Theta}_{t+1}' = \hat{\Theta}_t + \frac{(y_{t+1} - \hat{\Theta}_t\phi_t)\phi_t^\top}{N_t}, \quad N_t = \max\{\mu_0, \|\phi_t\|^2\}
\end{equation}
for any $\mu_0 > 0$, where the projection in \eqref{eqn:adaptive_law} is with respect to the Frobenius norm. $\{Z_t\}_{t \geq 1}$ is a sequence of compact, convex subsets of $\bbR^{m \times (n+m)}$. For now, we will suppose that $\{Z_t\}_{t \geq 1}$ satisfies the following condition:
\begin{condition} \label{cond:theta_in_set}
    For all $t \geq 1$, $\Theta_* \in Z_t \subseteq S_\Theta$.
\end{condition}
\noindent From Assumption \ref{asn:params_bounded}, a na{\"i}ve choice satisfying Condition \ref{cond:theta_in_set} would be $Z_t = S_\Theta\ \forall t \geq 1$, where
\begin{equation} \label{eqn:S_theta}
    S_\Theta = \{\Theta = [\Theta_A, \Theta_B] \in \bbR^{m \times (n+m)} : \Theta_A \in S_A \land \Theta_B \in S_B\}.
\end{equation}
In Algorithm \ref{alg:mrac-lqr}, we will make a more sophisticated choice of $Z_t$ which will lead to the optimality result in Theorem \ref{thm:regret}.

Combining \eqref{eqn:plant}, \eqref{eqn:control_input}, and \eqref{eqn:online_regression_definitions} and defining $\tilde{\Theta}_t = \hat{\Theta}_t - \Theta_*$, we see that the closed-loop plant satisfies
\begin{equation} \label{eqn:closed_loop_plant}
    x_{t+1} = A_{mk_t}x_t + B_m(r_t - \tilde{\Theta}_t\phi_t) + w_{t+1}.
\end{equation}
Then, combining \eqref{eqn:comparator_system} and \eqref{eqn:closed_loop_plant} and defining $e_t = x_t - x_{ct}$, we see that the error between the plant and comparator system is related to the parameter error by the following error model:
\begin{equation} \label{eqn:error_model}
    e_{t+1} = A_{mk_t}e_t - B_m\tilde{\Theta}_t\phi_t.
\end{equation}

Our first main result is a stability guarantee for the proposed adaptive controller, which holds for any bounded $\{r_t\}_{t \geq 0}$ regardless of excitation. This stability guarantee is summarized in the following two results, which are presented separately here for use later on. Theorem \ref{thm:stability} is proven in Appendix \ref{app:stability}. The proof of Proposition \ref{prop:regressor_stability} largely follows Appendices A.2, B.1.1, and B.1.2 of \cite{fisher2023MastersThesis} with slight adjustments to accommodate the time-varying reference dynamics, and thus a proof sketch is given in Appendix \ref{app:regressor_stability}, and the reader is referred to \cite{fisher2023MastersThesis} for further technical details.
\begin{proposition} \label{prop:regressor_stability}
    Let Assumptions \ref{asn:matching_condition} and \ref{asn:params_bounded} and Conditions \ref{cond:TVRM} and \ref{cond:theta_in_set} hold. Then, for the closed-loop adaptive system consisting of \eqref{eqn:plant}, \eqref{eqn:comparator_system}, and \eqref{eqn:control_input}-\eqref{eqn:adaptive_law} and $\phi_t$ as defined in \eqref{eqn:online_regression_definitions}, $\exists \lambda_2 \in (0, 1)$ and $d_0, d_1, d_2, d_3 > 0$ such that $\|\phi_t\|$ is a sub-Gaussian random variable with
    \begin{equation} \label{eqn:phi_t_subG_norm}
        \|\|\phi_t\|\|_{\psi_2} \leq d_0\lambda_2^t\|\phi_0\| + \frac{1 - \lambda_2^t}{1 - \lambda_2}\left(d_1 + d_2\sqrt{n}\sigma_w + d_3\sqrt{m}\sigma_\eta\right)
    \end{equation}
    for all $t \geq 0$.
\end{proposition}
\begin{theorem} \label{thm:stability}
    Let Assumptions \ref{asn:matching_condition} and \ref{asn:params_bounded} and Conditions \ref{cond:TVRM} and \ref{cond:theta_in_set} hold. Then, for the closed-loop adaptive system consisting of \eqref{eqn:plant}, \eqref{eqn:comparator_system}, and \eqref{eqn:control_input}-\eqref{eqn:adaptive_law}, $\|e_{ct}\|$ is a sub-Gaussian random variable
    whose sub-Gaussian norm is uniformly bounded. Additionally, the following implication holds with probability one:
    \begin{align}
        \sum_{t=0}^{T-1} \|\tilde{\Theta}_t\phi_t\|^2 = \tilde{O}(\sqrt{T}) &\implies \sum_{t=1}^T \|e_{ct}\|^2 = \tilde{O}(\sqrt{T}). \label{eqn:thetatilde_to_e_relation_2}
    \end{align}
\end{theorem}

%% file: Sections/Algorithm.tex
\section{Adaptive LQR algorithm and discussion} \label{sec:algorithm}

\begin{algorithm}[h]
    \caption{MRAC-LQR: adaptive LQR using direct MRAC with reference model updates}
    \label{alg:mrac-lqr}
    \begin{algorithmic}[1]
        \STATE{{\bf Require:} Initial reference model $(A_m, B_m)$; Initial state $x_0$; Parameter set $S_\Theta$; Initial parameter estimate $\hat{\Theta}_0 \in S_\Theta$; Cost matrices $Q, R$; Frequency set $\{\omega_i\}_{i=1}^N$ with $N \geq \frac{n+m}{2}$; Amplitude set $\{\bar{r}_i\}_{i=1}^N$, $\bar{r}_i \in \bbR^m$; Constants $C_T, C_\Lambda, \lambda > 0$, $\delta \in (0, 1)$}
        \STATE{{\bf Initialize} $\Xi_0 = \hat{\Theta}_0$, $\Lambda_0 = \Sigma_0^{-1} = \lambda I_{n+m}$, $A_{m0} = A_m$, $\Delta_{m0} = 0$, $t_0 = 0$, $k = 0$}
        \FOR{$t = 0, 1, 2, \dots$}
            \STATE{$r_t \gets (k+1)^{-1/2}\sum_{i=1}^N \bar{r}_i\sin(\omega_it)$} \COMMENT{Exploratory input} \label{algln:r_t}
            \STATE{$[\hat{\Theta}_{At}, \hat{\Theta}_{Bt}] \gets \hat{\Theta}_t\ \mbox{s.t.}\ \hat{\Theta}_{At} \in \bbR^{m \times n}, \hat{\Theta}_{Bt} \in \bbR^{m \times m}$}
            \STATE{$u_t \gets \mbox{Equation \eqref{eqn:control_input}}$} \COMMENT{Control input} \label{algln:control_input}
            \STATE{$\begin{cases} \mbox{{\bf Apply} control input}\ u_t \\
            \mbox{{\bf Measure} new state}\ x_{t+1} \end{cases}$} \COMMENT{Interact with the dynamical system for one time step}
            \STATE{$\phi_t, y_{t+1} \gets \mbox{Equation \eqref{eqn:online_regression_definitions}}$}
            \STATE{$\Xi_{t+1}, \Sigma_{t+1} \gets \mbox{Equations \eqref{eqn:RLS_covariance_update}-\eqref{eqn:RLS_param_update}}$} \COMMENT{RLS update}
            \STATE{$\calC_\delta(\Xi_{t+1}, \Sigma_{t+1}) \gets \mbox{Equation \eqref{eqn:C_delta}}$}
            \STATE{$Z_{t+1} \gets S_\Theta \cap \calC_\delta(\Xi_{t+1}, \Sigma_{t+1})$} \COMMENT{High-confidence parameter set} \label{algln:parameter_set}
            \STATE{$\hat{\Theta}_{t+1} \gets \mbox{Equation \eqref{eqn:adaptive_law}}$} \COMMENT{Adaptive law} \label{algln:adaptive_law}
            \IF{$\lambda_{min}(\Sigma_{t+1}^{-1} - \Lambda_k) \geq C_\Lambda$ \AND $t+1 - t_k \geq C_T(k+1)$} \label{algln:ref_model_update_conditions}
                \STATE{$\hat{A}_{k+1} \gets A_m - B_m\hat{\Theta}_{At_{k+1}}$, $\hat{B}_{k+1} \gets B_m\hat{\Theta}_{Bt_{k+1}}$} \COMMENT{Estimate the dynamics} \label{algln:dynamics_est}
                \STATE{$\hat{K}_{k+1} \gets \mathrm{dlqr}(\hat{A}_{k+1}, \hat{B}_{k+1}, Q, R)$} \COMMENT{Estimate the optimal gain} \label{algln:dlqr}
                \STATE{$A_{m(k+1)} \gets \hat{A}_{k+1} + \hat{B}_{k+1}\hat{K}_{k+1}$} \COMMENT{Update the reference dynamics} \label{algln:reference_model_update}
                \STATE{$\Delta_{m(k+1)} \gets \hat{\Theta}_{Bt_{k+1}}\hat{K}_{k+1} - \hat{\Theta}_{At_{k+1}}$} \COMMENT{Update the gain offset} \label{algln:gain_offset} 
                \STATE{$\begin{cases} \Lambda_{k+1} \gets \Sigma_{t+1}^{-1} \\ t_{k+1} \gets t+1 \\ k \gets k + 1 \end{cases}$} \COMMENT{Move on to next epoch} \label{algln:next_epoch}
            \ENDIF
        \ENDFOR
    \end{algorithmic}
\end{algorithm}

Our approach, MRAC-LQR, is summarized in the pseudocode in Algorithm \ref{alg:mrac-lqr}.
The adaptive control design proposed in Section \ref{subsec:adaptive_control_design} is reiterated in Lines \ref{algln:r_t}-\ref{algln:adaptive_law}, and constitutes a fast inner control loop.
Meanwhile, the reference model update in Lines \ref{algln:dynamics_est}-\ref{algln:next_epoch} constitutes a slow outer loop which seeks to optimize the reference model, using information learned from the exploratory signal $r_t$ in Line \ref{algln:r_t} to ensure $A_{mk} \to A_* + B_*K_*$. The shift in the reference model from $A_m$ to $A_{mk}$ appears in the control input as an extra gain $\Delta_{mk}$. Finally, to prevent persistent extra cost from exploration, $\|r_t\|$ decays with a carefully-chosen rate whose significance is made clear in the next section.

Note that our adaptive control design maintains two parameter estimates, $\Xi_t$ and $\hat{\Theta}_t$. $\Xi_t$ is generated from RLS and is used to produce sets $Z_t$ in Line \ref{algln:parameter_set} which contain $\Theta_*$ with high probability. These sets shrink as information accumulates in the RLS covariance matrix $\Sigma_t$. Meanwhile, $\hat{\Theta}_t$ constantly seeks values within $Z_t$ which stabilize the closed-loop dynamics. Intuitively, $\hat{\Theta}_t$ provides stability, while $\Xi_t$ constrains $\hat{\Theta}_t$ to ensure convergence to $\Theta_*$ under persistent excitation.

The success of Algorithm \ref{alg:mrac-lqr} relies on carefully choosing when to close the outer loop and update the reference model. The RLS covariance matrix $\Sigma_t$ stores the total information accumulated from $\{\phi_t\}_{t \geq 0}$, $\Lambda_k$ stores the information accumulated up to the start of epoch $k$, and $t_k$ is the first time step contained in epoch $k$. The transition from one epoch to the next occurs when the two conditions in Line \ref{algln:ref_model_update_conditions} are met. The condition on $\Sigma_t$ prevents the magnitude of $r_t$ from decaying before we are able to learn from it, while the condition on $t$ prevents the reference model from varying too rapidly and is essential for the stability guarantees in Theorems \ref{thm:TVRM_stability} and \ref{thm:stability}.

One key point to note is that, of the four categories presented in Section \ref{sec:related_work}, MRAC-LQR lies in the CE with Attenuating Exploration category. It may be possible to remove the exploration and replace it with an OFU-type approach. However, this would eliminate one of MRAC-LQR's benefits, which is computational efficiency. Additionally, parameter learning through exploration is beneficial for reasons other than convergence to optimality: for example, for preventing bursting \citep{annaswamy2023arcra}.

Also of note is that, in Line \ref{algln:r_t}, we propose a fixed sinusoidal exploration along the lines of \citep{sarker2023accurate}, rather than the Gaussian noise exploration typically found in related work. This is because, as discussed by \cite{sarker2023accurate}, real dynamical systems always have unmodeled dynamics, which are typically higher-frequency phenomena. In order to avoid exciting unmodeled dynamics, which effectively act as a state-dependent disturbance, it is essential to be able to limit the bandwidth of the input. For a linear time-invariant system, a sum of sufficiently many sinusoids is sufficient for exploration, while allowing for better control of the bandwidth than noise-based exploration.


A final key point, and one of the key benefits of MRAC-LQR, is the fact that the exploration in line \ref{algln:r_t} is needed \emph{only for optimality, not for stability}, as proven in Theorems \ref{thm:TVRM_stability} and \ref{thm:stability}. In the event that current control objectives do not permit exploration, $r_t$ can simply be set to zero at any time, even at the beginning, without loss of closed-loop stability.

%% file: Sections/Analysis.tex
\section{Analysis of Algorithm \ref{alg:mrac-lqr}} \label{sec:analysis}

First, we show that Algorithm \ref{alg:mrac-lqr} is globally stable with no requirements on an initial stabilizing controller or exploration online. As the stability guarantees in Section \ref{sec:adaptive_control} are dependent on Conditions \ref{cond:TVRM} and \ref{cond:theta_in_set}, we accomplish this by showing that Algorithm \ref{alg:mrac-lqr} satisfies these two conditions. This is done in the following proposition, proven in Appendix \ref{app:conditions_and_stability}:
\begin{proposition} \label{prop:conditions_and_stability}
    Let $r_t$ in Line \ref{algln:r_t} in Algorithm \ref{alg:mrac-lqr} be replaced by any uniformly bounded $\{r_t\}_{t \geq 0}$. Then, Algorithm \ref{alg:mrac-lqr} satisfies Condition \ref{cond:TVRM}, and, denoting the event on which Algorithm \ref{alg:mrac-lqr} satisfies Condition \ref{cond:theta_in_set} as $\calE_2$, $\calE_2$ holds with probability at least $1 - \delta$. Furthermore, on the event $\calE_2$, Algorithm \ref{alg:mrac-lqr} satisfies $\lim_{T \to \infty} \frac{1}{T}\sum_{t=1}^T \|e_{ct}\|^2 = 0$ with probability one.
\end{proposition}

We now present the main result of this work, which is a high-probability regret bound in the presence of exploration, proven in Appendix \ref{app:regret}:
\begin{theorem} \label{thm:regret}
    Consider the plant in \eqref{eqn:plant} subject to Algorithm \ref{alg:mrac-lqr}. With probability at least $1 - \delta - \delta'$ for any $\delta' \in (0, 1)$, we have $\Regret(T) \leq \tilde{O}(\sqrt{T})$.
\end{theorem}
While formal proofs are relegated to the appendices, we provide an intuition here. First, we combine Propositions \ref{prop:spectral_line_from_input}, \ref{prop:FE_from_spectral_lines}, \ref{prop:confidence_set_convergence}, and \ref{prop:regressor_stability} and Theorem \ref{thm:stability} to show that, with high probability and neglecting polylog factors, $T_k$ grows linearly with $k$. Additionally, it follows from Proposition \ref{prop:confidence_set_convergence} and Line \ref{algln:ref_model_update_conditions} of Algorithm \ref{alg:mrac-lqr} that, again neglecting polylog terms, for any $k \geq 0$ and any $t \in [t_k, t_{k+1})$, $\|\tilde{\Theta}_t\|_F^2$ decreases as $(k+1)^{-1}$, which is the same rate at which $\|r_t\|^2$ decreases.

Next, defining $\tilde{\Theta}_{At} = \hat{\Theta}_{At} - \Theta_{A*}$ and $\tilde{\Theta}_{Bt} = \hat{\Theta}_{Bt} - \Theta_{B*}$, \eqref{eqn:control_input} can be rewritten via straightforward algebra as
\begin{equation} \label{eqn:control_input_rewritten}
    u_t = \bar{K}_tx_t + \hat{\Theta}_{Bt}^{-1}r_t, \quad \bar{K}_t = \hat{K}_{k_t} + \hat{\Theta}_{Bt}^{-1}((\tilde{\Theta}_{Bt_{k_t}} - \tilde{\Theta}_{Bt})\hat{K}_{k_t} + \tilde{\Theta}_{At} - \tilde{\Theta}_{At_{k_t}}).
\end{equation}
Then, the regret is decomposed into three terms: $\Regret(T) = R_1(T) + R_2(T) + R_3(T)$, where $R_1(T)$ is regret due to convergence of $\bar{K}_t$ to $K_*$ in \eqref{eqn:opt_K}, $R_2(T)$ is additional cost incurred by the exploratory signal $\hat{\Theta}_{Bt}^{-1}r_t$, and $R_3(T)$ consists of cross terms. For bounding $R_1(T)$, note that $\bar{K}_t$ is the sum of two gains: $\hat{K}_{k_t}$, which is the optimal gain matrix for $(\hat{A}_{k_t}, \hat{B}_{k_t})$, and another gain which has magnitude $O(\|\tilde{\Theta}_t\| + \|\tilde{\Theta}_{t_{k_t}}\|)$. Thus, using results from \cite{simchowitz2020naive} on the LQR costs of feedback gains which are close to $K_*$, as well as a Lyapunov-like analysis to handle the time-varying nature of $\bar{K}_t$ and convert from expected cost to a high-probability bound, we achieve $R_1(T) \leq \tilde{O}(\sqrt{T})$. Bounding $R_2(T)$ is easier: it is a sum of terms which are all $O(\|r_t\|^2)$, and we straightforwardly obtain $R_2(T) \leq \tilde{O}(\sqrt{T})$ using $\|r_t\|^2 = O((k_t+1)^{-1})$ and $T_k = \tilde{O}(k+1)$. Finally, care must be taken in bounding $R_3(T)$, as the continuously time-varying nature of $\bar{K}_t$ makes statistical analysis challenging. However, by carefully decomposing the terms in $R_3(T)$, we obtain a sum which can be bounded by the sub-Gaussian Hoeffding inequality \citep{Vershynin2019}, plus a sum of terms which are all $\tilde{O}(\|\tilde{\Theta}_t\|\|r_t\|)$. The aforementioned rates of decay for $\|\tilde{\Theta}_t\|$ and $\|r_t\|$, together with $T_k = \tilde{O}(k+1)$, lead to $R_3(T) \leq \tilde{O}(\sqrt{T})$.

%% file: Sections/Simulations.tex
\section{Simulations} \label{sec:simulations}


We now present simulation results comparing MRAC-LQR to other certainty equivalence methods.
In what follows, the baseline optimal controller is $u_t = K_*x_t$ with no adaptation or excitation; nominal CE is the naive approach in \citep{mania2019CEefficient}; CE based on FIR truncated SLS is the approach in \citep{Dean_2018}; and MRAC-LQR is Algorithm \ref{alg:mrac-lqr}. All simulations use $Q = 10I$ and $R = I$, and all algorithms use linear epoch scheduling instead of exponential, as is typical in the simulations of related works. Additionally, in order to make fair comparisons, all algorithms use the same type of exploration - either Gaussian or deterministic - and all algorithms are modified to take advantage of Assumption \ref{asn:matching_condition} to prevent MRAC-LQR from having an unfair advantage due to estimating fewer parameters. See Appendix \ref{app:PE_sims} for additional simulation results with deterministic sinusoidal exploration.

\begin{figure}[h]
    \centering
    \begin{subfigure}[t]{0.49\textwidth}
        \centering
        \caption{Regret}
        \label{fig:laplacian_stable_explore_0.1_regret}
        \includegraphics[width=\textwidth]{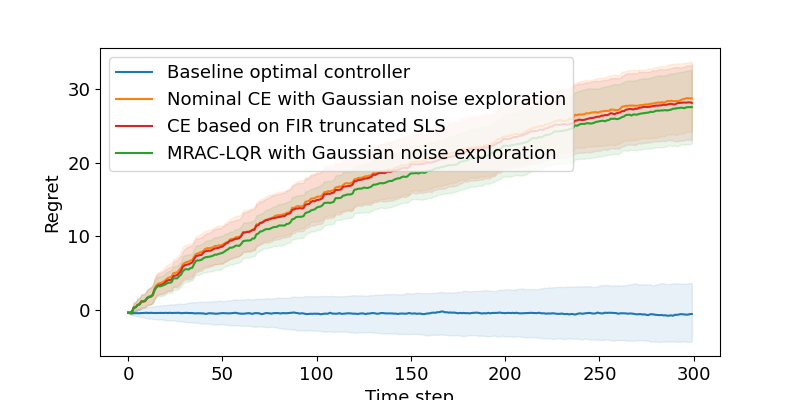}
    \end{subfigure}
    \hfill
    \begin{subfigure}[t]{0.49\textwidth}
        \centering
        \caption{State magnitude}
        \label{fig:laplacian_stable_explore_0.1_state}
        \includegraphics[width=\textwidth]{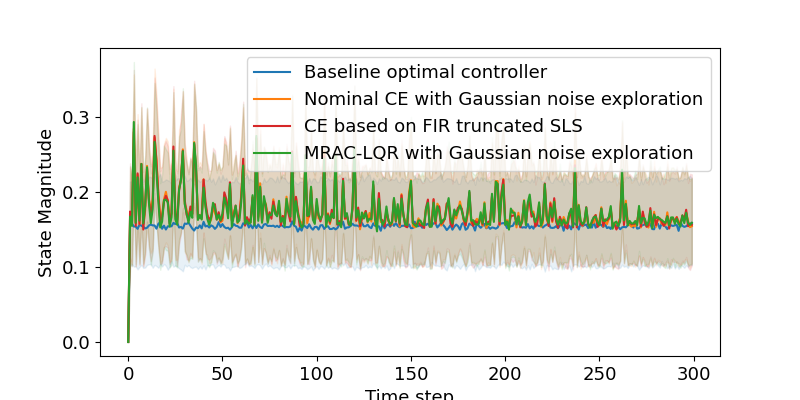}
    \end{subfigure}
    \caption{Laplacian system with stable initial controller and $\sigma_{\rm explore} = 0.1$. Solid lines are the median values over 1000 trials, and shaded regions are the 20\%-80\% confidence windows.}
    \label{fig:laplacian_stable_explore_0.1}
\end{figure}
\begin{figure}[h]
    \centering
    \begin{subfigure}[t]{0.49\textwidth}
        \centering
        \caption{Regret}
        \label{fig:laplacian_unstable_explore_0.1_regret}
        \includegraphics[width=\textwidth]{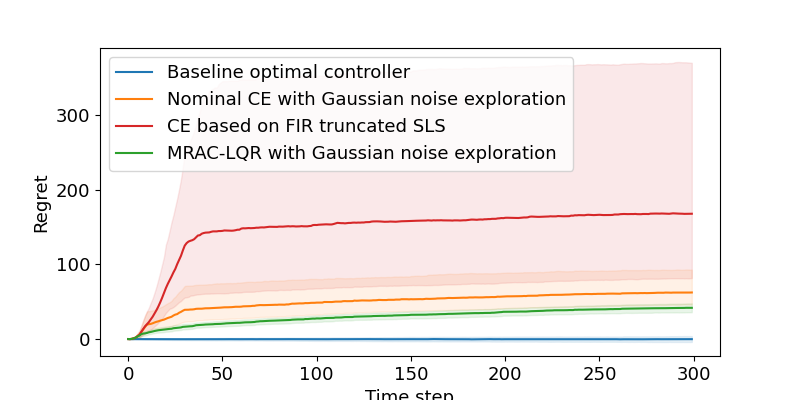}
    \end{subfigure}
    \hfill
    \begin{subfigure}[t]{0.49\textwidth}
        \centering
        \caption{State magnitude}
        \label{fig:laplacian_unstable_explore_0.1_state}
        \includegraphics[width=\textwidth]{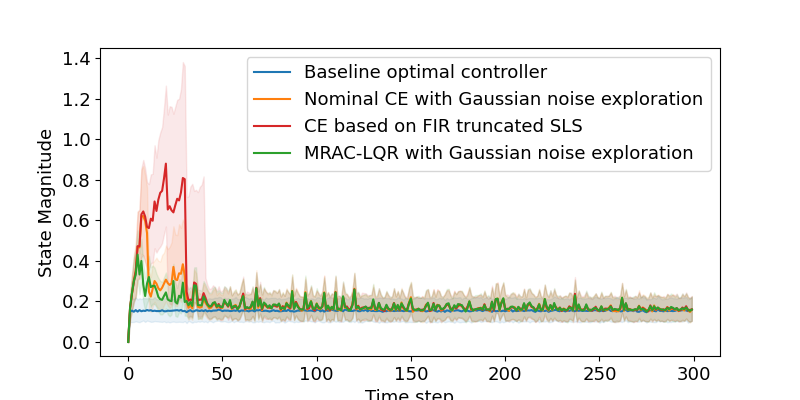}
    \end{subfigure}
    \caption{Laplacian system with unstable initial controller and $\sigma_{\rm explore} = 0.1$. Solid lines are the median values over 1000 trials, and shaded regions are the 20\%-80\% confidence windows.}
    \label{fig:laplacian_unstable_explore_0.1}
\end{figure}

Appendices \ref{app:gaussian_sims} and \ref{app:PE_sims} provide simulation details and additional simulation results for two dynamical systems: a marginally unstable Laplacian system as in \citep{Dean_2018}, and a quadrotor linearized about hover with partial loss of rotor effectiveness. See Appendix \ref{app:sim_dynamics} for the plant dynamics. Here, we show simulation results for the Laplacian system under Gaussian noise exploration of substantial magnitude. In Figure \ref{fig:laplacian_stable_explore_0.1}, we see that when an initial stabilizing controller is given and the magnitude of the exploratory signal is significant - ideal conditions for indirect adaptive control methods - MRAC-LQR still performs just as well as the nominal CE method.
It should be noted, however, that MRAC-LQR performs much better when the parametric uncertainty is larger (see Figure \ref{fig:laplacian_unstable_explore_0.1}). In Figure \ref{fig:laplacian_unstable_explore_0.1}, the magnitude of the initial uncertainty is such that the initial feedback gains are not stabilizing for the true plant. The other methods eventually stabilize the system, but suffer large initial costs from applying an unstable controller. MRAC-LQR, in contrast, quickly stabilizes the closed-loop system, keeping the state magnitude smaller and the cost lower.

%% file: Sections/Related_Work.tex
\section{Related work} \label{sec:related_work}

The following brief literature survey divides the most relevant adaptive LQR work into two categories.
These categories are not necessarily all-encompassing, but are meant to help the reader in gaining an understanding of the literature. Other literature categories that may be of interest, although less relevant to the current work, are reinforcement-learning based adaptive LQR approaches \citep{bradtke1994adaptiveLQR,jiang2012adaptiveoptimal,fazel2018policygradient,mohammadi2021dlqr}, iterative feedback tuning \citep{hjalmarsson1998ift}, and data-driven control techniques \citep{pang2018datadriven,berberich2020datadriven,vanwaarde2020datainformativity,dorfler2023datadriven}.

\subsubsection*{Certainty equivalence with attenuating exploration}

Certainty Equivalence (CE) refers to any adaptive algorithm in which the current best parameter estimate is used in lieu of the true parameters in control design. This category has generally focused on indirect adaptive controllers: least squares is used to estimate the plant parameters, and the current best estimate is periodically used to calculate a new optimal feedback gain. As it is well-known that indirect adaptive LQR can be unstable unless the parameters converge asymptotically to their true values, these approaches also inject an exploratory signal. However, exploration adds to the LQR cost online, so to prevent the cost from being permanently suboptimal, these approaches let the magnitude of the exploration go to zero with carefully-chosen rates of decay.

It was shown in the 1980s \citep{chen1986leastsquares_lowercase} that this approach can lead to convergence in expectation of the online cost to its optimal value. In recent years, regret bounds have become a popular way to quantify the rate of cost convergence. \cite{Dean_2018} derived the first regret bound for the certainty equivalence approach, at $\tilde{O}(T^{2/3})$. This bound was subsequently strengthened to $\tilde{O}(\sqrt{T})$ by \cite{mania2019CEefficient}, and following work by \cite{simchowitz2020naive} proved that certainty equivalence achieves both the best possible dependence on $T$ and near-optimal dependence on system dimension. Finally, while much of the adaptive LQR literature uses Gaussian noise as the exploratory signal, it was shown by \cite{sarker2023accurate} that a deterministic sum of sinusoids works just as well and is more robust to unmodeled dynamics.

Algorithms in this category are computationally efficient, and regrets have good dependence on both time and system dimension. However, as prior work in this category is based on indirect adaptive control, existing approaches require either an initial stabilizing policy or extremely small initial parameter errors. MRAC-LQR also falls into this category, but our leveraging of direct adaptive control allows us to avoid this requirement. In lieu of an initial stabilizing controller, we require Assumption \ref{asn:matching_condition}, which asks for a dynamically-feasible stable system which the adaptive controller can attempt to emulate.

\subsubsection*{Cost-biased/optimism in the face of uncertainty}

The cost-biased, or Optimism in the Face of Uncertainty (OFU), approach is another indirect adaptive control approach. The original goal of this approach was to obtain convergence to optimality online without requiring an extra exploratory signal. The original approach in \citep{Campi98} added a term to the least-squares cost function that favored parameter estimates with lower LQR costs over estimates with higher costs, hence "cost-biased." This approach was proven to result asymptotic cost convergence online, at the cost of needing to solve a non-convex optimization at every other time step.

Subsequently, with a reformulation and a name change in \citep{abbasi2011}, the cost-biased approach became OFU. The regret bound obtained by \cite{abbasi2011} was the first $\tilde{O}(\sqrt{T})$ regret bound obtained for the adaptive LQR problem. However, OFU still required periodically solving a constrained non-convex optimization problem, and its regret scaled exponentially with system dimension. Subsequent work in \citep{ibrahimi2012sparseadaptiveLQR} resolved the poor dimension dependence for the special case of sparse dynamical systems. More recently, in \citep{cohen2019SDPdare}, a SDP relaxation of the DARE was developed, leading to an OFU algorithm with polynomial dependence on system dimension (although the exponent could be as large as $n^{16}$ per \citealp{simchowitz2020naive}). However, this approach also relies on either an initial stabilizing controller or an extremely small initial parameter error. Finally, it is worth noting that although OFU is stable in theory, the state magnitude can become very large in practice. A recent algorithm, StabL, proposed by \cite{lale2022reinforcement}, addressed this issue by injecting an exploratory signal of constant magnitude over a finite time window.

In summary, cost-biased/OFU has the benefit of not requiring exploration for closed-loop stability, at least in theory. However, it is computationally very expensive, and recent work has shown that the need for exploration may be inescapable in practice.

%% file: Sections/Future_Work.tex
\section{Future work} \label{sec:future_work}

This work has a number of future directions to take. We elucidate several below, in order from the most obvious to the least immediately apparent (but perhaps most interesting).

\subsection*{Assumption \ref{asn:matching_condition}}

It would be desirable to eliminate, or at least relax, Assumption \ref{asn:matching_condition}. This is in general not easy to do while maintaining an excitation-independent stability guarantee. One potential option is adaptive pole placement control (APPC) (see e.g., \citealp{guo1996WLSAdaptiveControl,Landau11}), in which the reference model update would be replaced by an update to the desired poles.

\subsection*{Linear-Quadratic-Gaussian (LQG)}


In a practical sense, adaptive LQG is an even more important problem than adaptive LQR, as real applications almost never have the full state available for measurement.
Future work may extend our approach to adaptive LQG in the case where cost function only depends on the measurements $\{y_t\}_{t \geq 0}$ by designing an adaptive controller using output feedback techniques for direct MRAC (see \citealp{Goodwin_1984}).

\subsection*{The case where $Q$ and $R$ depend on $A_*$ and $B_*$}

This work, and most of the adaptive LQR literature, seeks to optimize with respect to fixed cost matrices $Q$ and $R$ chosen a priori. In practice, though, when designing an LQR controller, $Q$ and $R$ are chosen according to a number of factors, including control objectives, state and input constraints, and the plant dynamics. Consider an example where the uncertainty $\Theta_{B*}$ is due to a damaged actuator with an unknown loss of effectiveness coefficient: if the actuator damage were known ahead of time, one would choose $R$ to penalize use of the damaged actuator. Thus, there is an ideal $R_*$, which is a function of the unknown parameters.

Algorithm \ref{alg:mrac-lqr} may be straightforwardly extended to a general case where $(Q_*, R_*) = f(A_*, B_*)$ for some function $f$ as follows: add the line $(\hat{Q}_{k+1}, \hat{R}_{k+1}) \gets f(\hat{A}_{k+1}, \hat{B}_{k+1})$ between Lines \ref{algln:dynamics_est} and \ref{algln:dlqr}, and change Line \ref{algln:dlqr} to use $\hat{Q}_{k+1}$ and $\hat{R}_{k+1}$.

\subsection*{Adaptive LQR with time-varying $A_*$ and $B_*$}

All prior works cited in Section \ref{sec:related_work} that solve the adaptive LQR problem - including OFU, as shown by \cite{lale2022reinforcement} - rely on parameter learning to ensure closed-loop stability. As a result, it is impossible to introduce any significant forgetting factor into prior approaches without compromising stability. MRAC-LQR has a global stability guarantee which does not require $\hat{\Theta}_t \to \Theta_*$, and is thus uniquely able to accommodate a forgetting factor in the RLS parameter estimator in \eqref{eqn:RLS_cov_inv_update}-\eqref{eqn:RLS_param_update}. This enables us in future work to consider a problem extension of high practical importance: the setting where $A_*$ and $B_*$ may vary unpredictably with time, corresponding to real-time environment changes or damage.

%% file: Sections/Conclusions.tex
\section{Conclusions} \label{sec:conclusions}

This work proposes a new framework for addressing the adaptive LQR problem: a fast inner control loop to track a stable reference model using direct adaptive control methods, inside a slower outer loop to apply exploration, learn the parameters, and update the reference model to converge to optimality. This framework is made concrete in a new algorithm, MRAC-LQR, which is proven to stabilize an LTI system with unknown parameters and with no requirements on open-loop stability, initial parameter error, or excitation. This is in contrast to existing methods, which obtain stability only through either (i) an initial stabilizing controller, (ii) parameter learning via exploration, and/or (iii) computationally intensive algorithms. Additionally, when exploration is added, MRAC-LQR is proven to attain $\Regret(T) \leq \tilde{O}(\sqrt{T})$. Simulation results show that MRAC-LQR performs as well as the best existing algorithm when an initial stabilizing controller is known and when exploration is applied. Additionally, when either the initial controller is not stabilizing or the exploration magnitude is set very small, MRAC-LQR significantly outperforms existing methods due to its ability to stabilize quickly. Finally, several extensions of MRAC-LQR were proposed for future work which extend the adapt-stabilize-learn-optimize framework to related problems of interest.

%% file: Appendices/Random_Variables.tex
\section{} \label{app:preliminaries}

This appendix provides supporting material for Section \ref{sec:preliminaries}, and additional preliminaries that will be used in subsequent analysis.

\subsection{Useful properties of random distributions} \label{app:random_variables}

Here, we provide a few useful properties of sub-Gaussian and sub-exponential random variables. Results from other works are presented without proof.


\begin{lemma}[\citealp{Vershynin2019}] \label{lem:subG_norm}
    $\|\cdot\|_{\psi_2}$ is a norm on the space of sub-Gaussian random variables, and thus satisfies $\|a_1X_1 + a_2X_2\|_{\psi_2} \leq |a_1|\|X_1\|_{\psi_2} + |a_2|\|X_2\|_{\psi_2}$ for any sub-Gaussian random variables $X_1, X_2$ and any constants $a_1, a_2$.
\end{lemma}

\begin{lemma}[\citealp{Vershynin2019}] \label{lem:subG_tail_bound}
    A sub-Gaussian random variable $X$ satisfies $\bbP[|X| \geq c] \leq 2\exp(-C_2c^2/\|X\|_{\psi_2}^2)\ \forall c \geq 0$ for a universal constant $C_2 > 0$.
\end{lemma}

\begin{lemma}[\citealp{Vershynin2019}] \label{lem:ZMsubG_is_subG}
    A random variable $X \sim \ZMsubG(\sigma^2)$ has sub-Gaussian norm $\|X\|_{\psi_2} = C_2'\sigma$ for a universal constant $C_2' > 0$.
\end{lemma}

\begin{lemma}[\citealp{Vershynin2019}] \label{lem:subE_norm}
    $\|\cdot\|_{\psi_1}$ is a norm on the space of sub-exponential random variables, and thus satisfies $\|a_1X_1 + a_2X_2\|_{\psi_1} \leq |a_1|\|X_1\|_{\psi_1} + |a_2|\|X_2\|_{\psi_1}$ for any sub-exponential random variables $X_1, X_2$ and any constants $a_1, a_2$.
\end{lemma}

\begin{lemma}[\citealp{Vershynin2019}] \label{lem:subE_tail_bound}
    A sub-exponential random variable $X$ satisfies $\bbP[|X| \geq c] \leq 2\exp(-C_1c/\|X\|_{\psi_1})\ \forall c \geq 0$ for a universal constant $C_1 > 0$.
\end{lemma}

\begin{lemma}[\citealp{Vershynin2019}] \label{lem:subE_is_subG_squared}
    A random variable $X$ is sub-Gaussian iff $X^2$ is sub-exponential. Furthermore, $\|X^2\|_{\psi_1} = \|X\|_{\psi_2}^2$.
\end{lemma}

\begin{proposition}[\citealp{Vershynin2019}] \label{prop:Hoeffding}
    Let $X_1, \dots, X_N$ be independent zero-mean random variables, and let $a_1, \dots, a_N$ be real constants. Then, for every $c \geq 0$, we have
    \begin{equation*}
        \bbP\left[\left|\sum_{i=1}^N a_iX_i\right| \geq c\right] \leq 2\exp\left(-\frac{C_2''c^2}{\max_i\|X_i\|_{\psi_2}^2\sum_{i=1}^N a_i^2}\right)
    \end{equation*}
    for a universal constant $C_2'' > 0$.
\end{proposition}

\begin{lemma} \label{lem:norm_of_subG_vector}
    Consider a real-valued random vector $X \sim \ZMsubG(\sigma^2I_d)$. Then, $\|X\|^2$ is sub-exponential with $\|\|X\|^2\|_{\psi_1} \leq O(d\sigma^2)$ and $\|X\|$ is sub-Gaussian with $\|\|X\|\|_{\psi_2} \leq O(\sqrt{d}\sigma)$.
\end{lemma}
\proof{
    Denote $X_i$ as the $i$th element of $X$. From Definition \ref{def:ZMsubG_vector}, we know that $X_i \sim \ZMsubG(\sigma^2)\ \forall i$. Then, using the lemmas above, $\|X_i\|_{\psi_2} = C_2'\sigma \implies \|X_i^2\|_{\psi_1} = (C_2'\sigma)^2 \implies \|\|X\|^2\|_{\psi_1} = \|\sum_{i=1}^d X_i^2\|_{\psi_1} \leq d(C_2'\sigma)^2 \implies \|\|X\|\|_{\psi_2} \leq \sqrt{d}C_2'\sigma$.
\qed}

\subsection{Stability of slowly time-varying linear systems}

The following result was introduced by \cite{Kreisselmeier1986} to prove stability for adaptive control with slowly time-varying parameters, and was adapted by \cite{Miller2017} to prove stability for adaptive control under exogenous disturbances. Here, we provide a special case which is sufficient for our purposes. This proposition will be useful in proving the stability results in Theorem \ref{thm:TVRM_stability} and Proposition \ref{prop:regressor_stability}.

\begin{proposition}[Adapted from \citealp{Kreisselmeier1986,Miller2017}] \label{prop:LTV_stability}
    Consider the dynamical system $x_{t+1} = (A_t + \Delta_t)x_t$ with state transition matrix $\Phi(t, \tau)$. Suppose that there exist constants $\gamma_1 > 1$, $\lambda_A \in (0, 1)$, and $\alpha_1, \beta_1, \beta_2 \geq 0$ such that
    \begin{enumerate}
        \item for all $t \geq 0$, $\|A_t^i\|_\op \leq \gamma_1\lambda_A^i\ \forall i \geq 0$,
        \item for all $t \geq \tau \geq 0$, $\sum_{s=\tau}^{t-1} \|A_{t+1} - A_t\|_\op \leq \alpha_1\sqrt{t - \tau}$,
        \item for all $t \geq \tau \geq 0$, $\sum_{s=\tau}^{t-1} \|\Delta_t\|_\op \leq \beta_1\sqrt{t - \tau} + \beta_2(t - \tau)$, and
        \item there exist constants $\bar{\lambda} \in (\lambda_A, 1)$ and $N \in \bbN$ satisfying $\beta_2 < \frac{1}{\gamma_1}(\frac{\bar{\lambda}}{\gamma_1^{1/N}} - \lambda_A)$.
    \end{enumerate}
    Then, there exists a constant $\gamma_2$ such that $\|\Phi(t, \tau)\|_\op \leq \gamma_2\bar{\lambda}^{t - \tau}\ \forall t \geq \tau \geq 0$.
\end{proposition}

\subsection{Stability and optimality of nearby feedback gains}

The following results are adapted from \cite{simchowitz2020naive}, and will be useful in proving our regret bound in Theorem \ref{thm:regret}. Given any stabilizable dynamics pair $(A, B)$ and any gain matrix $K \in \bbR^{m \times n}$ which stabilizes $(A, B)$, define $P_{\infty(A, B)}[K]$ as the solution to
\begin{equation} \label{eqn:dlyap_cost}
    (A + BK)^\top P_{\infty(A, B)}[K](A + BK) - P_{\infty(A, B)}[K] = -(Q + K^\top RK).
\end{equation}
As we assume that $Q > 0$, we have $P_{\infty(A, B)}[K] = P_{\infty(A, B)}[K]^\top > 0$. Additionally, from \eqref{eqn:opt_K}-\eqref{eqn:opt_dare}, one can show that $P_{\infty(A_*, B_*)}[K_*] = P_*$, which we will make use of later on.

Now, let $(\hat{A}, \hat{B})$ be any estimate of $(A_*, B_*)$, and define $\hat{K} = \dlqr(\hat{A}, \hat{B}, Q, R)$. Also define $P_{\rm lyap}$ as the symmetric positive-definite solution to
\begin{equation} \label{eqn:dlyap}
    (A_* + B_*K_*)^\top P_{\rm lyap}(A_* + B_*K_*) - P_{\rm lyap} = -I_n.
\end{equation}
The following results from \cite{simchowitz2020naive} relate the dynamics estimation error to the gain error, and allow us to quantify the stability and optimality of gains close to $K_*$.

\begin{proposition}[Adapted from \citealp{simchowitz2020naive}] \label{prop:dynamics_error_to_gain_error}
    For norms $\circ \in \{\op, F\}$, define the error $\epsilon_\circ = \max\{\|\hat{A} - A_*\|_\circ, \|\hat{B} - B_*\|_\circ\}$. Then, if $\epsilon_\op \leq \frac{1}{16\|P_*\|_\op^2}$, the dynamics pair $(\hat{A}, \hat{B})$ is stabilizable, and the following bounds hold:
    \begin{enumerate}
        \item $\|\sqrt{R}(\hat{K} - K_*)\|_\circ \leq 7(2\|P_*\|_\op^2)^{7/4}\epsilon_\circ$, and
        \item $\|B_*(\hat{K} - K_*)\|_\circ \leq 8(2\|P_*\|_\op^2)^{7/4}\epsilon_\circ$.
    \end{enumerate}
\end{proposition}

\begin{proposition}[Adapted from \citealp{simchowitz2020naive}] \label{prop:optimality_of_gain}
    Fix any gain matrix $K \in \bbR^{m \times n}$ satisfying $\|B_*(K - K_*)\|_\op \leq \frac{1}{5\|P_*\|_\op^{3/2}}$. Then, the following bounds hold:
    \begin{enumerate}
        \item $\|P_{\infty(A_*, B_*)}[K] - P_*\|_\op \leq \|P_*\|_\op\max\{\|K - K_*\|_\op^2, \|\sqrt{P_*}B_*(K - K_*)\|_\op^2\}$, and
        \item $(A_* + B_*K)^\top P_{\rm lyap}(A_* + B_*K) - P_{\rm lyap} \preceq -\frac{1}{2}\|P_{\rm lyap}\|_\op^{-1}P_{\rm lyap}$.
    \end{enumerate}
\end{proposition}

\subsection{Bound on sums of martingale difference sequences}

The following result from \cite{chen1991Identification} allows us to bound the sum of a martingale difference sequence with high probability. This result will be useful in proving our regret bound in Theorem \ref{thm:regret}.

\begin{proposition}[Adapted from \citealp{chen1991Identification}] \label{prop:mds_bound}
    Let $\{X_t, \calF_t\}$ be a matrix martingale difference sequence and $\{M_t, \calF_t\}$ be an adapted sequence of random matrices with $\|M_t\| < \infty\ a.s.\ \forall t \geq 0$ and $\sup_{t \geq 0} \bbE[\|X_{t+1}\|^2 | \calF_t] < \infty\ a.s.$. Define $s_t = \sqrt{\sum_{t=0}^T \|M_t\|^2}$. Then, as $T \to \infty$, we have $\sum_{t=0}^T M_tX_{t+1} = O(s_t\ln^{\eta + 1/2}(s_t^2 + e))\ a.s.\ \forall \eta > 0$.
\end{proposition}

\subsection{Proof of Proposition \ref{prop:confidence_set_convergence}} \label{app:confidence_set_convergence}

On the event $\calE_1$, we have
\begin{equation*}
    \Tr[(\Xi_t' - \Theta_*)\Sigma_t^{-1}(\Xi_t' - \Theta_*)^\top] \leq 4\beta_\delta(\Sigma_t)\ \forall t \geq 0.
\end{equation*}
Furthermore, we can bound $\det(\Sigma_t^{-1}) \leq \|\Sigma_t^{-1}\|_2^d \leq (\lambda + \sum_{s=0}^{t-1} \|\phi_s\|^2)^d$. Suppose that $\|\|\phi_t\|\|_{\psi_2} \leq \sigma_\phi\ \forall t \geq 0$, and define $z_t = \sum_{s=0}^{t-1} \|\phi_s\|^2$. Then, from Lemma \ref{lem:subE_is_subG_squared}, $z_t$ is a sub-exponential random variable with $\|z_t\|_{\psi_1} \leq t\sigma_\phi^2$. Applying Lemma \ref{lem:subE_tail_bound}, for any $\delta' \in (0, 1)$ and any $t \geq 0$, we have $z_t \leq \frac{t\sigma_\phi^2}{C_1}\ln(\frac{2}{\delta'})$ with probability at least $1 - \delta'$.
Defining $C' = \frac{\sigma_\phi^2}{C_1\lambda}\ln(\frac{2}{\delta'})$ for convenience, we thus have
\begin{equation} \label{eqn:Xi_prime_bound}
    \Tr[(\Xi_t' - \Theta_*)\Sigma_t^{-1}(\Xi_t' - \Theta_*)^\top] \leq O\left(\ln\left(\frac{1 + C't}{\delta^{2/d}}\right)\right).
\end{equation}

Now, consider any orthonormal basis vectors $\{q_i\}_{i=1}^d \in \bbR^d$. As the $q_i$ are orthonormal, we have $\sum_{i=1}^d q_iq_i^\top = I_d$, and thus we can bound
\begin{equation} \label{eqn:Xi_prime_phi_decomposition}
    \sum_{t=0}^{T-1} \|(\Xi_t' - \Theta_*)\phi_t\|^2 = \sum_{t=0}^{T-1} \left\|\sum_{i=1}^d (\Xi_t' - \Theta_*)q_iq_i^\top\phi_t\right\|^2 \leq d\sum_{i=1}^d \sum_{t=0}^{T-1} (\phi_t^\top q_i)^2\|(\Xi_t' - \Theta_*)q_i\|^2.
\end{equation}
It therefore suffices to consider the accumulated information along each direction $q_i$, and the resulting parameter error along each direction. The accumulated information along $q_i$ up to time $T$ is given by $\sum_{t=0}^{T-1} (\phi_s^\top q_i)^2$.
We will consider two possible cases for each $q_i$: either (1) $\sum_{t=0}^{T-1} (\phi_s^\top q_i)^2 \leq O(\ln(\frac{1 + C'T}{\delta^{2/d}})\sqrt{T})$, or (2) $\sum_{t=0}^{T-1} (\phi_s^\top q_i)^2 > O(\ln(\frac{1 + C'T}{\delta^{2/d}})\sqrt{T})$.

\subsubsection*{Case (1): $\sum_{t=0}^{T-1} (\phi_s^\top q_i)^2 \leq O(\ln(\frac{1 + C'T}{\delta^{2/d}})\sqrt{T})$}

In this case, using the fact that $\Xi_t' \in S_\Theta\ \forall t \geq 0$, we know that $\|(\Xi_t' - \Theta_*)q_i\|^2$ is uniformly bounded in $t$, and thus that
\begin{equation*}
    \sum_{t=0}^{T-1} (\phi_t^\top q_i)^2\|(\Xi_t' - \Theta_*)q_i\|^2 = O\left(\sum_{t=0}^{T-1} (\phi_t^\top q_i)^2\right) \leq O\left(\ln\left(\frac{1 + C'T}{\delta^{2/d}}\right)\sqrt{T}\right).
\end{equation*}

\subsubsection*{Case (2): $\sum_{t=0}^{T-1} (\phi_s^\top q_i)^2 > O(\ln(\frac{1 + C'T}{\delta^{2/d}})\sqrt{T})$}

In this case, decomposing the RLS covariance as $\Sigma_t^{-1} = \sum_{j=1}^d \sigma_{jt}q_jq_j^\top$, we must have $\sigma_{it} > O(\ln(\frac{1 + C't}{\delta^{2/d}})\sqrt{t})$. Therefore, from \eqref{eqn:Xi_prime_bound}, we have
\begin{gather*}
    \sigma_{it}\|(\Xi_t' - \Theta_*)q_i\|^2 \leq \Tr\left[\sum_{j=1}^d \sigma_{jt}(\Xi_t' - \Theta_*)q_jq_j^\top(\Xi_t' - \Theta_*)^\top\right] \leq O\left(\ln\left(\frac{1 + C't}{\delta^{2/d}}\right)\right) \\
    \implies \|(\Xi_t' - \Theta_*)q_i\|^2 < O\left(t^{-1/2}\right).
\end{gather*}
Furthermore, as $\|\phi_t\|$ is sub-Gaussian with $\|\|\phi_t\|\|_{\psi_2} \leq \sigma_\phi$, $\|\phi_t\|^2$ is sub-exponential with $\|\|\phi_t\|\|_{\psi_1} \leq \sigma_\phi^2$, and Lemma \ref{lem:subE_tail_bound} along with a union bound shows that for any $\delta'' \in (0, 1)$, $\max_{t \in [0, T)} \|\phi_t\|^2 \leq \frac{\sigma_\phi^2}{C_1}\ln(\frac{2T}{\delta''})$ with probability at least $1 - \delta''$. Thus,
\begin{equation*}
    \sum_{t=0}^{T-1} (\phi_t^\top q_i)^2\|(\Xi_t' - \Theta_*)q_i\|^2 \leq \left(\max_{t \in [0, T)} \|\phi_t\|^2\right)\sum_{t=0}^{T-1} \|(\Xi_t' - \Theta_*)q_i\|^2 \leq O\left(\ln\left(\frac{2T}{\delta''}\right)\sqrt{T}\right).
\end{equation*}

\subsubsection*{Completing the proof}

In both cases, we have $\sum_{t=0}^{T-1} (\phi_t^\top q_i)^2\|(\Xi_t' - \Theta_*)q_i\|^2 \leq \tilde{O}(\sqrt{T})$ for every $q_i$. The claim follows from \eqref{eqn:Xi_prime_phi_decomposition}.
\qed

%% file: Appendices/Comparator_System_Proofs.tex
\section{}

This appendix provides an analysis of the adaptive controller proposed in Section \ref{sec:adaptive_control}. First, we show that Condition \ref{cond:TVRM} leads to a uniform bound over $k$ on the magnitude of the state transition matrix of $A_{mk}$. This bound allows us to prove the comparator system stability result in Theorem \ref{thm:TVRM_stability}. Then, we provide proof sketches of Proposition \ref{prop:regressor_stability} and Theorem \ref{thm:stability}, and the reader is referred to \cite{fisher2023MastersThesis} for full technical details.

\subsection{Bound on the state transition matrix of $A_{mk_t}$}

\begin{proposition} \label{prop:A_m_state_transition_matrix}
    Consider the dynamical system $x_{t+1} = A_{mk_t}x_t$, with corresponding state transition matrix denoted as $\Phi_m(t, \tau)$. There exist constants $\gamma_1 > 1$ and $\lambda_1 \in (0, 1)$ such that $\|\Phi_m(t, \tau)\|_\op \leq \gamma_1\lambda_1^{t - \tau}\ \forall t \geq \tau \geq 0$.
\end{proposition}
\proof{
     $A_{mk_t}$ is constant with respect to $t$ except for the time steps when epoch transitions occur, and from Condition \ref{cond:TVRM}(i) and (ii), we know that the jumps at epoch transitions are bounded. In short, we have
    \begin{equation} \label{eqn:delta_A_m_bound}
        \|A_{mk_t} - A_{mk_{t-1}}\|_\op \leq \begin{cases} 0, & t \notin \{t_k\}_{k \geq 1} \\ 2\|B_m\|_\op\Delta_{max}, & t \in \{t_k\}_{k \geq 1} \end{cases}.
    \end{equation}
    It follows from \eqref{eqn:delta_A_m_bound} that, for any $t \geq \tau \geq 0$,
    \begin{equation} \label{eqn:sum_delta_A_m_bound}
        \sum_{s=\tau}^{t-1} \|A_{mk_{s+1}} - A_{mk_s}\|_\op \leq 2\|B_m\|_\op\Delta_{max}N_\tau^t
    \end{equation}
    where $N_\tau^t$ denotes the number of epoch transitions from time $\tau$ through time $t$. It is straightforward to see that, as Condition \ref{cond:TVRM}(iv) requires the length of each epoch to increase at least linearly, we have $N_\tau^t \leq O(\sqrt{t - \tau})$. Finally, Condition \ref{cond:TVRM}(iii) states that there exist $\gamma_m > 1$ and $\lambda_m \in (0, 1)$ such that $\|A_{mk_t}^i\|_\op \leq \gamma_m\lambda_m^i\ \forall t \geq 0, i \geq 0$. The claim follows from Proposition \ref{prop:LTV_stability} with $A_t = A_{mk_t}$, $\Delta_t = 0$, $\lambda_A = \lambda_m$, $\gamma_1 = \gamma_m$, and $\alpha_1 = O(\|B_m\|_\op\Delta_{max})$.
\qed}

\subsection{Bounding the regressor} \label{app:bounding_growth_rates}

Algorithm \ref{alg:mrac-lqr} uses the regressor $\phi_t = [-x_t^\top, u_t^\top]^\top$ throughout. For analysis, however, it is useful to define an auxiliary regressor $\xi_t = [x_t^\top, r_t^\top]^\top$. Here, we provide lemmas bounding the relative growth rates of the two regressors.

\begin{lemma} \label{lem:phi_O_xi}
    There exists a constant $D_1 > 0$ such that $\|\phi_t\|^2 \leq D_1\|\xi_t\|^2$.
\end{lemma}
\proof{
    Using the definitions of $\phi_t$ in \eqref{eqn:online_regression_definitions} and $u_t$ in \eqref{eqn:control_input}, we can write
    \begin{align*}
        \|\phi_t\|^2 &= \|x_t\|^2 + \|u_t\|^2 = \|x_t\|^2 + \|\hat{\Theta}_{Bt}^{-1}((\hat{\Theta}_{At} + \Delta_{mk_t})x_t + r_t)\|^2 \\
        &\leq \|x_t\|^2 + 2\|\hat{\Theta}_{Bt}^{-1}(\hat{\Theta}_{At} + \Delta_{mk_t})x_t\|^2 + 2\|\hat{\Theta}_{Bt}^{-1}r_t\|^2 \\
        &\leq (1 + 4\|\hat{\Theta}_{Bt}^{-1}\|_\op^2(\|\hat{\Theta}_{At}\|_\op^2 + \|\Delta_{mk_t}\|_\op^2))\|x_t\|^2 + 2\|\hat{\Theta}_{Bt}^{-1}\|_\op^2\|r_t\|^2 \\
        &\leq \|\xi_t\|^2\sup_{t \geq 0}\max\{1 + 4\|\hat{\Theta}_{Bt}^{-1}\|_\op^2(\|\hat{\Theta}_{At}\|_\op^2 + \|\Delta_{mk_t}\|_\op^2), 2\|\hat{\Theta}_{Bt}^{-1}\|_\op^2\}.
    \end{align*}
    Condition \ref{cond:TVRM}(ii) states that $\|\Delta_{mk_t}\|_\op \leq \Delta_{max}$, and the adaptive law in \eqref{eqn:adaptive_law} along with Condition \ref{cond:theta_in_set} guarantees that $\hat{\Theta}_t \in S_\Theta\ \forall t \geq 0$. Then, from Assumption \ref{asn:params_bounded}, we know that $\|\hat{\Theta}_{At}\|_\op$ and $\|\hat{\Theta}_{Bt}^{-1}\|_\op$ are bounded. The claim follows.
\qed}

\begin{lemma} \label{lem:xi_O_phi}
    There exists a constant $D_2 > 0$ such that $\|\xi_t\|^2 \leq D_2\|\phi_t\|^2$.
\end{lemma}
\proof{
    First, note that \eqref{eqn:control_input} can be rewritten as
    \begin{equation*}
        r_t = \hat{\Theta}_{Bt}u_t - (\hat{\Theta}_{At} + \Delta_{mk_t})x_t.
    \end{equation*}
    Then, we can write
    \begin{align*}
        \|\xi_t\|^2 &= \|x_t\|^2 + \|r_t\|^2 = \|x_t\|^2 + \|\hat{\Theta}_{Bt}u_t - (\hat{\Theta}_{At} + \Delta_{mk_t})x_t\|^2 \\
        &\leq \|x_t\|^2 + 2\|(\hat{\Theta}_{At} + \Delta_{mk_t})x_t\|^2 + 2\|\hat{\Theta}_{Bt}u_t\|^2 \\
        &\leq (1 + 4(\|\hat{\Theta}_{At}\|_\op^2 + \|\Delta_{mk_t}\|_\op^2))\|x_t\|^2 + 2\|\hat{\Theta}_{Bt}\|_\op^2\|u_t\|^2 \\
        &\leq \|\phi_t\|^2\sup_{t \geq 0} \max\{1 + 4(\|\hat{\Theta}_{At}\|_\op^2 + \|\Delta_{mk_t}\|_\op^2), 2\|\hat{\Theta}_{Bt}\|_\op^2\}.
    \end{align*}
    The claim follows similarly to the proof of Lemma \ref{lem:phi_O_xi}.
\qed}

\subsection{Bounding the state of an exponentially stable system}

The following lemma bounds the sum over time of a dynamical system's state by the sum over time of its control inputs, which will be useful in proving Theorem \ref{thm:stability}.

\begin{lemma} \label{lem:sums_of_sequences}
    Consider two sequences $\{a_t\}_{t \geq 0}$, $\{b_t\}_{t \geq 0}$ such that $a_t \leq \sum_{s=0}^{t-1} \lambda^{t-s-1}b_s\ \forall t \geq 1$ for some $\lambda \in (0, 1)$. Then, for any $T \geq 1$, we have $\sum_{t=1}^T a_t \leq \frac{1}{1 - \lambda}\sum_{t=0}^{T-1} b_t$.
\end{lemma}
\proof{
We use the fact that $\sum_{i=0}^{N-1} \lambda^i = \frac{1 - \lambda^N}{1 - \lambda}$ for any $N \geq 1$, $\lambda \in (0, 1)$. Expanding and regrouping sums, we obtain
\begin{align*}
    \sum_{t=1}^T a_t &\leq \sum_{t=1}^T\sum_{s=0}^{t-1} \lambda^{t-s-1}b_s = b_0\sum_{i=0}^{T-1} \lambda^i + b_1\sum_{i=0}^{T-2} \lambda^i + \cdots + b_{T-2}\sum_{i=0}^1 \lambda^i + b_{T-1} \\
    &= \sum_{t=0}^{T-1} b_t\frac{1 - \lambda^{T-t}}{1 - \lambda} \leq \frac{1}{1 - \lambda}\sum_{t=0}^{T-1} b_t.
\end{align*}
\qed}

\subsection{Proof of Theorem \ref{thm:TVRM_stability}} \label{app:TVRM_stability}

Using the bound on the state transition matrix $\Phi_m$ derived in Proposition \ref{prop:A_m_state_transition_matrix}, we can immediately write and bound $x_{ct}$ for any $t \geq 0$ as (see e.g., \citealp{hespanha2018LinearSystemsTheory})
\begin{gather}
    x_{ct} = \Phi(t, 0)x_0 + \sum_{s=0}^{t-1} \Phi_m(t, s+1)(B_mr_s + w_{s+1}) \implies \label{eqn:x_ct} \\
    \|x_{ct}\| \leq \gamma_1\lambda_1^t\|x_0\| + \gamma_1\sum_{s=0}^{t-1} \lambda_1^{t-s-1}(\|B_m\|_\op\|r_s\| + \|w_{s+1}\|). \label{eqn:x_ct_bound}
\end{gather}
From \eqref{eqn:subG_norm}, we see that any deterministic $a \in \bbR$ is sub-Gaussian with $\|a\|_{\psi_2} = \frac{|a|}{\sqrt{\ln(2)}}$. Furthermore, from Lemma \ref{lem:norm_of_subG_vector}, we have $\|\|w_t\|\|_{\psi_2} \leq O(\sqrt{n}\sigma_w)\ \forall t \geq 1$. Finally, \eqref{eqn:x_ct_subG_norm} follows from Lemma \ref{lem:subG_norm} and the fact that $\sum_{s=0}^{t-1} \lambda_1^{t-s-1} = \sum_{i=0}^{t-1} \lambda_1^i = \frac{1 - \lambda_1^t}{1 - \lambda_1}$.
\qed

\subsection{Proof of Proposition \ref{prop:regressor_stability} (sketch)} \label{app:regressor_stability}

First, note that \eqref{eqn:adaptive_law} can be rewritten using \eqref{eqn:online_regression} as
\begin{equation} \label{eqn:adaptive_law_rewritten}
    \hat{\Theta}_{t+1} = \proj_{Z_{t+1}}[\hat{\Theta}_{t+1}'], \quad \hat{\Theta}_{t+1}' - \Theta_* = \tilde{\Theta}_t - \frac{\tilde{\Theta}_t\phi_t\phi_t^\top - \eta_{t+1}\phi_t^\top}{N_t}.
\end{equation}
By convexity of $Z_t$, we can write $\|\proj_{Z_{t+1}}[\hat{\Theta}_{t+1}'] - \Theta_*\|_F \leq \|\hat{\Theta}_{t+1}' - \Theta_*\|_F$. Now, define a sequence $V_t = \|\tilde{\Theta}_t\|_F^2$. Using \eqref{eqn:adaptive_law} and simplifying, we obtain (\citealp{fisher2023MastersThesis}, Appendix B.1.1)
\begin{equation} \label{eqn:V_increment}
    V_{t+1} - V_t \leq -\frac{\|\tilde{\Theta}_t\phi_t\|^2}{N_t} + 2\frac{\|\tilde{\Theta}_t\phi_t\|}{\sqrt{N_t}}\frac{\|\eta_{t+1}\|}{\sqrt{N_t}} + \frac{\|\eta_{t+1}\|^2}{N_t}.
\end{equation}
Now, fix any $\epsilon > 0$, and consider any two time steps $\tau, t \in \bbZ_{\geq 0}$ with $\tau < t$ such that $\frac{\|\eta_{s+1}\|^2}{N_s} \leq \epsilon\ \forall s \in [\tau, t)$. From Assumption \ref{asn:params_bounded}, the fact that $\hat{\Theta}_t \in S_\Theta\ \forall t \geq 0$, and the fact that $m < n$ by Assumption \ref{asn:matching_condition} and thus that $\rank(\tilde{\Theta}_{At}), \rank(\tilde{\Theta}_{Bt}) \leq m$, we have $0 \leq V_t \leq 4m(a_{max}^2 + b_{max}^2)\ \forall t \geq 0$. Using \eqref{eqn:V_increment} and these bounds on $V_t$ and simplifying, we obtain (\citealp{fisher2023MastersThesis}, Appendix B.1.2)
\begin{equation} \label{eqn:adaptive_law_decay_rate_when_noise_small}
    \sum_{s=\tau}^{t-1} \frac{\|\tilde{\Theta}_s\phi_s\|}{\sqrt{N_s}} \leq 2\sqrt{m(a_{max}^2 + b_{max}^2)(t - \tau)} + \epsilon(1 + \sqrt{2})(t - \tau).
\end{equation}

Now, using \eqref{eqn:closed_loop_plant}, the dynamics of the auxiliary regressor $\xi_t$ can be written as
\begin{equation} \label{eqn:xi_dynamics}
    \xi_{t+1} = \underbrace{\begin{bmatrix} A_{mk_t} & B_m \\ 0 & 0 \end{bmatrix}}_{\bar{A}_{k_t}}\xi_t + \underbrace{\begin{bmatrix} I_n \\ 0 \end{bmatrix}}_{\bar{B}_1}(w_{t+1} - B_m\tilde{\Theta}_t\phi_t) + \underbrace{\begin{bmatrix} 0 \\ I_m \end{bmatrix}}_{\bar{B}_2}r_{t+1}.
\end{equation}
Define $\calN_t = \max\{\mu_0, \|\xi_t\|^2\}$, $\bar{\Delta}_t = \frac{1}{\calN_t}\bar{B}_1B_m\tilde{\Theta}_t\phi_t\xi_t^\top$ and $\nu_t = (1 - \frac{\|\xi_t\|^2}{\calN_t})\tilde{\Theta}_t\phi_t$, and note that using Lemmas \ref{lem:phi_O_xi} and \ref{lem:xi_O_phi}, we have $N_t \leq D_1^2\calN_t$, $\calN_t \leq D_2^2N_t$, and $\|\nu_t\| \leq \frac{4D_1}{3}\sqrt{\frac{\mu_0(a_{max}^2 + b_{max}^2)}{3}} \\ \forall t \geq 0$. Equation \eqref{eqn:xi_dynamics} can then be rewritten as
\begin{equation} \label{eqn:xi_dynamics_rewritten}
    \xi_{t+1} = (\bar{A}_{k_t} - \bar{\Delta}_t)\xi_t + \bar{B}_1(w_{t+1} - B_m\nu_t) + \bar{B}_2r_{t+1}.
\end{equation}

We now examine the state transition matrix $\Phi_\xi(t, \tau)$ associated with the time-varying dynamics matrix $\bar{A}_{k_t} - \bar{\Delta}_t$ over any set of consecutive time steps on which the noise-to-signal ratio is small: specifically, over any time window $[\tau, t)$ on which $\frac{\|\eta_{s+1}\|^2}{N_s} \leq \epsilon\ \forall s \in [\tau, t)$ for an $\epsilon > 0$ to be chosen later. For any $t \geq 0$ and $i \geq 1$, Condition \ref{cond:TVRM}(iii) guarantees
\begin{gather*}
    \bar{A}_{k_t}^i = \begin{bmatrix} A_{mk_t}^i & A_{mk_t}^{i-1}B_m \\ 0 & 0 \end{bmatrix} \implies \\
    \|\bar{A}_{k_t}^i\|_\op \leq \|A_{mk_t}^i\|_\op + \|A_{mk_t}^{i-1}B_m\|_\op \leq \gamma_m\lambda_m^i + \|B_m\|_\op\gamma_m\lambda_m^{i-1} = \gamma_m\left(1 + \frac{\|B_m\|_\op}{\lambda_m}\right)\lambda_m^i.
\end{gather*}
Also, as in the proof of Proposition \ref{prop:A_m_state_transition_matrix}, we have
\begin{equation*} \label{eqn:sum_delta_A_bar_bound}
    \sum_{s=\tau}^{t-1} \|\bar{A}_{k_{s+1}} - \bar{A}_{k_s}\|_\op \leq 2\|B_m\|_\op\Delta_{max}N_\tau^t
\end{equation*}
where $N_\tau^t \leq O(\sqrt{t - \tau})$. Additionally, as $\frac{\|\eta_{s+1}\|^2}{N_s} \leq \epsilon\ \forall s \in [\tau, t)$, using \eqref{eqn:adaptive_law_decay_rate_when_noise_small}, we have
\begin{align*}
    \sum_{s=\tau}^{t-1} \|\bar{\Delta}_s\|_\op &\leq \sum_{s=\tau}^{t-1} \frac{\|B_m\|_\op\|\tilde{\Theta}_s\phi_s\|\|\xi_s\|}{\calN_s} \leq \sum_{s=\tau}^{t-1} \frac{D_1\|B_m\|_\op\|\tilde{\Theta}_s\phi_s\|}{\sqrt{N_s}} \\
    &\leq D_1\|B_m\|_\op\left(2\sqrt{m(a_{max}^2 + b_{max}^2)(t - \tau)} + \epsilon(1 + \sqrt{2})(t - \tau)\right).
\end{align*}
Finally, we can invoke Proposition \ref{prop:LTV_stability} with $A_t = \bar{A}_{k_t}$, $\Delta_t = \bar{\Delta}_t$, $\lambda_A = \lambda_m$, $\gamma_1 = \gamma_m(1 + \frac{\|B_m\|_\op}{\lambda_m})$, $\alpha_1 = O(\|B_m\|_\op\Delta_{max})$, $\beta_1 = 2D_1\|B_m\|_\op\sqrt{m(a_{max}^2 + b_{max}^2)}$, and $\beta_2 = \epsilon(1 + \sqrt{2})D_1\|B_m\|_\op$, choosing $\epsilon$ sufficiently small that condition (iii) in the proposition is satisfied, to conclude that there exist $\gamma_2 > 1$, $\lambda_2 \in (0, 1)$ such that $\|\Phi_\xi(s_2, s_1)\|_\op \leq \gamma_2\lambda_2^{s_2 - s_1}\ \forall \tau \leq s_1 \leq s_2 < t$.

The proof of Proposition \ref{prop:regressor_stability} concludes by partitioning the set of time steps $\bbZ_{\geq 0}$ into a series of time intervals which alternate between low noise-to-signal ratio (i.e. $\frac{\|\eta_{t+1}\|^2}{N_t} \leq \epsilon$ over the interval) and high noise-to-signal ratio (i.e. $\frac{\|\eta_{t+1}\|^2}{N_t} > \epsilon$ over the interval). Over time intervals with low noise-to-signal ratio, we bound $\|\xi_t\|$ using the bound $\|\Phi_\xi(t, \tau)\|_\op \leq \gamma_2\lambda_2^{t - \tau}$ and the bound on $\|\nu_t\|$. Over time intervals with high noise-to-signal ratio, $\|\xi_t\|$ is straightforwardly bounded by the magnitude of the noise. Finally, we combine these bounds into one bound on $\|\xi_t\|$ which covers all time steps, and \eqref{eqn:phi_t_subG_norm} follows from this bound and Lemma \ref{lem:phi_O_xi}. See Appendix A.2 of \cite{fisher2023MastersThesis} for details.
\qed

\subsection{Proof of Theorem \ref{thm:stability}} \label{app:stability}

From \eqref{eqn:error_model}, Proposition \ref{prop:A_m_state_transition_matrix}, and the initial condition $e_{c0} = 0$, we have
\begin{equation} \label{eqn:e_c_bound_1}
    \|e_{ct}\| \leq \gamma_1\|B_m\|_\op\sum_{s=0}^{t-1} \lambda_1^{t-s-1}\|\tilde{\Theta}_s\phi_s\|.
\end{equation}
A crude bound on the sub-Gaussian norm of $\|e_{ct}\|$ can be immediately established from \eqref{eqn:e_c_bound_1}, \eqref{eqn:phi_t_subG_norm}, and the fact that $\|\tilde{\Theta}_t\|_\op$ is uniformly bounded by projection, and it is easy to thus verify that $\|\|e_{ct}\|\|_{\psi_2}$ is uniformly bounded.

To prove that \eqref{eqn:thetatilde_to_e_relation_2} holds, define a sequence $E_t = e_{ct}^\top P_{mk_t}e_{ct}$, where $P_{mk_t}$ is the symmetric positive definite solution to the discrete-time Lyapunov equation
\begin{equation}
    A_{mk_t}^\top P_{mk_t}A_{mk_t} - P_{mk_t} = -I_n.
\end{equation}
Condition \ref{cond:TVRM} ensures that $P_{mk_t}$ is uniformly bounded with $\|P_{mk_t}\|_\op \leq \bar{P}\ \forall t \geq 0$ for a finite $\bar{P} > 1$. Additionally, define $\Delta P_{mt} = P_{mk_{t+1}} - P_{mk_t}$, and note that $\Delta P_{mt} = 0$ for all $t$ except where $k_{t+1} \neq k_t$. Then, using \eqref{eqn:error_model}, we obtain
\begin{align*}
    E_{t+1} - E_t &= -\|e_{ct}\|^2 - 2e_{ct}^\top A_{mk_t}^\top P_{mk_t}B_m\tilde{\Theta}_t\phi_t + \phi_t^\top\tilde{\Theta}_t^\top B_m^\top P_{mk_t}B_m\tilde{\Theta}_t\phi_t + e_{c(t+1)}^\top\Delta P_{mt}e_{c(t+1)} \\
    &\leq -\frac{\|e_{ct}\|^2}{2} + \phi_t^\top\tilde{\Theta}_t^\top B_m^\top(P_{mk_t} + 2P_{mk_t}A_{mk_t}A_{mk_t}^\top P_{mk_t})B_m\tilde{\Theta}_t\phi_t + e_{c(t+1)}^\top\Delta P_{mt}e_{c(t+1)} \\
    &\leq -\frac{E_t}{2\bar{P}} + \phi_t^\top\tilde{\Theta}_t^\top B_m^\top(P_{mk_t} + 2P_{mk_t}A_{mk_t}A_{mk_t}^\top P_{mk_t})B_m\tilde{\Theta}_t\phi_t + e_{c(t+1)}^\top\Delta P_{mt}e_{c(t+1)}
\end{align*}
where the first inequality uses the relation $-ax + 2b\sqrt{x} + c \leq -\frac{a}{2}x + \frac{2b^2}{a} + c$ for any $x, a, b, c > 0$. Using $E_0 = 0$ and summing from 0 to $t$, we obtain
\begin{equation} \label{eqn:E_bound}
    E_t \leq \sum_{s=0}^{t-1} \left(1 - \frac{1}{2\bar{P}}\right)^{t-s-1}\left(\bar{C}\|\tilde{\Theta}_s\phi_s\|^2 + e_{c(s+1)}^\top\Delta P_{ms}e_{c(s+1)}\right)
\end{equation}
for a system-dependent $\bar{C} > 0$. From \eqref{eqn:E_bound} and Lemma \ref{lem:sums_of_sequences} with $a_t = E_t$, $b_t = \bar{C}\|\tilde{\Theta}_t\phi_t\|^2 + e_{c(t+1)}^\top\Delta P_{mt}e_{c(t+1)}$, and $\lambda = 1 - \frac{1}{2\bar{P}}$, we finally obtain
\begin{equation*}
    \sum_{t=1}^T \|e_{ct}\|^2 = O\left(\sum_{t=1}^T E_t\right) \leq O\left(\sum_{t=0}^{T-1} \|\tilde{\Theta}_t\phi_t\|^2 + e_{c(t+1)}^\top\Delta P_{mt}e_{c(t+1)}\right).
\end{equation*}
$\|\Delta P_{mt}\|_\op \leq 2\bar{P}\ \forall t \geq 0$, and from Condition \ref{cond:TVRM}(iv), $\Delta P_{mt} = 0$ for all but at most $O(\sqrt{T})$ time steps. Furthermore, let $\|\|e_{ct}\|\|_{\psi_2} \leq \sigma_e\ \forall t \geq 0$. Then, $\|\|e_{ct}\|^2\|_{\psi_1} \leq \sigma_e^2$, and Lemma \ref{lem:subE_tail_bound} with a union bound shows that for any $\delta' \in (0, 1)$, $\max_{t \in [1, T]} \|e_{ct}\|^2 \leq \frac{\sigma_e^2}{C_1}\ln(\frac{2T}{\delta'})$ with probability at least $1 - \delta'$. Thus, $\sum_{t=0}^{T-1} e_{c(t+1)}^\top\Delta P_{mt}e_{c(t+1)} \leq O(\ln(\frac{2T}{\delta'})\sqrt{T})$, and \eqref{eqn:thetatilde_to_e_relation_2} follows.
\qed

%% file: Appendices/Regret_Proof_new.tex
\section{} \label{app:regret_proof}

This appendix provides an analysis of Algorithm \ref{alg:mrac-lqr}. First, we prove that all parameter, dynamics, and gain estimates are uniformly bounded. With this proof, then, we show that Algorithm \ref{alg:mrac-lqr} has the stability guarantees in Theorems \ref{thm:TVRM_stability} and \ref{thm:stability} with high probability by proving Proposition \ref{prop:conditions_and_stability}. Finally, we prove our main regret bound in Theorem \ref{thm:regret}.

\subsection{Boundedness of all gains} \label{app:gains_bounded}

The following lemma shows that all estimates and gains resulting from Algorithm \ref{alg:mrac-lqr} are uniformly bounded in time.

\begin{lemma} \label{lem:gains_bounded}
    Under Algorithm \ref{alg:mrac-lqr}, there exist constants $a_{max}, b_{max}, b_{min}, k_{max}, \Delta_{max} > 0$ such that $\|\hat{\Theta}_{At}\|_\op \leq a_{max}$, $\|\hat{\Theta}_{Bt}\|_\op \leq b_{max}$, and $\|\hat{\Theta}_{Bt}^{-1}\|_\op \leq \frac{1}{b_{min}}\ \forall t \geq 0$, and $\|\hat{K}_k\|_\op \leq k_{max}$ and $\|\Delta_{mk}\|_\op \leq \Delta_{max}\ \forall k \geq 0$.
\end{lemma}
\proof{
    The bounds on $\hat{\Theta}_{At}$ and $\hat{\Theta}_{Bt}$ follow immediately from Assumption \ref{asn:params_bounded}, the fact that $Z_t \subseteq S_\Theta\ \forall t \geq 1$, and the projection in \eqref{eqn:adaptive_law}. Now consider the estimated LQR covariance matrix $\hat{P}_k = \dare(\hat{A}_k, \hat{B}_k, Q, R)$. From Assumption \ref{asn:params_bounded}, the fact that $\hat{\Theta}_{At} \in S_A, \hat{\Theta}_{Bt} \in S_B\ \forall t \geq 0$, and $\hat{A}_k$ and $\hat{B}_k$ in Line \ref{algln:dynamics_est} of Algorithm \ref{alg:mrac-lqr}, we know that all dynamics estimates $(\hat{A}_k, \hat{B}_k)\ \forall k \geq 0$ lie in a compact set in which every element is stabilizable, and because we assume $Q > 0$, we know that $(\hat{A}_k, \sqrt{Q})$ is observable $\forall k \geq 0$. Therefore, all solutions $\hat{P}_k$ lie in a compact set, and we have $\|\hat{P}_k\|_\op \leq P_{max}\ \forall k \geq 0$ for some $P_{max} > 0$. Furthermore, it is well-known that $\hat{P}_k \geq Q\ \forall k \geq 0$ (see e.g. \citealp{lee1998DAREBounds}), and positive-definiteness of $Q$ thus gives us a uniform lower bound on $\hat{P}_k$. As $\hat{A}_k$, $\hat{B}_k$, and $\hat{P}_k$ are all bounded uniformly in $k$, we know that $\hat{K}_k$ and therefore $\Delta_{mk}$ are uniformly bounded as well.

    Finally, fix any $k \geq 0$, and consider the dynamical system $x_{t+1} = A_{mk}x_t$. Immediately, for any $x_0 \in \bbR^n$, we obtain
    \begin{equation*} \label{eqn:A_mk_bound_1}
        x_t = A_{mk}^tx_0\ \forall t \geq 0.
    \end{equation*}
    Furthermore, define a sequence $X_t = x_t^\top\hat{P}_kx_t$ for the same fixed value of $k$. One can show that the increment satisfies
    \begin{equation*}
        X_{t+1} - X_t = -x_t^\top(Q + \hat{K}_k^\top R\hat{K}_k)x_t \leq -\lambda_{min}(Q)\|x_t\|^2 \leq -\frac{\lambda_{min}(Q)}{P_{max}}X_t
    \end{equation*}
    and therefore that
    \begin{equation*} \label{eqn:A_mk_bound_2}
        X_t \leq \left(1 - \frac{\lambda_{min}(Q)}{P_{max}}\right)^tX_0 \implies \|x_t\|^2 \leq \frac{P_{max}}{\lambda_{min}(Q)}\left(1 - \frac{\lambda_{min}(Q)}{P_{max}}\right)^t\|x_0\|^2.
    \end{equation*}
    Finally, we conclude
    \begin{equation*}
        \|x_t\| = \|A_{mk}^tx_0\| \leq \sqrt{\frac{P_{max}}{\lambda_{min}(Q)}\left(1 - \frac{\lambda_{min}(Q)}{P_{max}}\right)^t}\|x_0\|\ \forall x_0 \in \bbR^n
    \end{equation*}
    and therefore $\|A_{mk}^i\|_\op \leq \gamma_m\lambda_m^i$ with $\gamma_m = \sqrt{\frac{P_{max}}{\lambda_{min}(Q)}}$, $\lambda_m = \sqrt{1 - \frac{\lambda_{min}(Q)}{P_{max}}}$.
\qed}

\subsection{Bounding the epoch lengths}

In this section and the next, we provide preliminary lemmas which will help in proving our regret bound in Theorem \ref{thm:regret}. To begin, we can rewrite the comparator system in \eqref{eqn:comparator_system} as
\begin{equation} \label{eqn:comparator_system_rewritten}
    x_{c(t+1)} = A_*x_{ct} + B_*v_t + w_{t+1}, \quad v_t = \Theta_{B*}^{-1}((\Theta_{A*} + \Delta_{mk_t})x_{ct} + r_t)
\end{equation}
Now, note that from Line \ref{algln:r_t} in Algorithm \ref{alg:mrac-lqr}, during epoch $k$, $\{r_t\}_{t \geq 0}$ has at least $n+m$ sub-Gaussian spectral lines from time $t_k$ to time $t_{k+1}-1$ with frequencies $\omega_i$, amplitudes $(k+1)^{-1/2}\bar{r}(\omega_i)$ for some constant linearly independent $\bar{r}(\omega_i) \in \bbC^m$, and radii $0$. Define the comparator regressor $\phi_{ct} := [-x_{ct}^\top, v_t^\top]^\top$. The following result characterizes the frequency content of the comparator system:
\begin{lemma} \label{lem:comparator_spectral_lines}
    During epoch $k$, $\{\phi_{ct}\} _{t \geq 0}$ has at least $n+m$ sub-Gaussian spectral lines from time $t_k$ to time $t_{k+1}-1$ with frequencies $\omega_i$, linearly independent amplitudes
    \begin{equation} \label{eqn:phi_c_amplitudes}
        \bar{\phi}_{ck}(\omega_i) = (k+1)^{-1/2}\begin{bmatrix} -(e^{j\omega_i}I_n - A_{mk})^{-1}B_m \\ \Theta_{B*}^{-1}(\Theta_{A*} + \Delta_{mk})(e^{j\omega_i}I_n - A_{mk})^{-1}B_m + \Theta_{B*}^{-1} \end{bmatrix}\bar{r}(\omega_i),
    \end{equation}
    and radii
    \begin{equation} \label{eqn:phi_c_radii}
        \sigma_{ck}(\omega_i) = \max\{1, \|\Theta_{B*}^{-1}(\Theta_{A*} + \Delta_{mk})\|_\op\}\|(e^{j\omega_i}I_n - A_{mk})^{-1}\|_\op\sigma_w.
    \end{equation}
\end{lemma}
\proof{Follows immediately from Proposition \ref{prop:spectral_line_from_input} and \eqref{eqn:comparator_system_rewritten}.
\qed}

With the frequency content of the comparator system characterized, its information content follows from Definition \ref{def:expected_info_matrix} and Proposition \ref{prop:FE_from_spectral_lines}. Then, the following result uses Lemma \ref{lem:comparator_spectral_lines} to bound the rate at which epoch lengths can grow.
\begin{lemma} \label{lem:epochs_linear}
    On the event $\calE_2$, we have $T_k \leq \tilde{O}(k+1)$ and $k_T \leq \tilde{O}(\sqrt{T})$ with probability one.
\end{lemma}
\proof{
First, denoting the expected information matrix of the comparator system during epoch $k$ as $\bar{\Phi}_{ck}$, from Lemma \ref{lem:comparator_spectral_lines}, \eqref{eqn:expected-info_matrix}, and \eqref{eqn:phi_c_amplitudes}, we have $\|\bar{\Phi}_{ck}^{-1}\|_\op = \bar{C}_{ck}\sqrt{k+1}$ for some $\bar{C}_{ck} > 0$. Define $\bar{\sigma}_{ck} := \max_i \sigma_{ck}(\omega_i)$, where $\sigma_{ck}(\omega_i)$ is given in \eqref{eqn:phi_c_radii}. It is easy to see from Lemma \ref{lem:gains_bounded} and \eqref{eqn:phi_c_radii} that there exist constants $\bar{C}_c > 0$ and $\bar{\sigma}_c > 0$ such that $\bar{C}_{ck} \leq \bar{C}_c$ and $\bar{\sigma}_{ck} \leq \bar{\sigma}_c\ \forall k \geq 0$. Choose any $\delta' \in (0, 1)$, define $\delta_k = \frac{6\delta'}{\pi^2(k+1)^2}$, and suppose that
\begin{equation} \label{eqn:T_k_lower_bound}
    T_k \geq \frac{2(n+m)\bar{\sigma}_c^2\ln(9^{2(n+m)}/\delta_k)\bar{C}_c^2(k+1)}{c} \geq \frac{2(n+m)\bar{\sigma}_{ck}^2\ln(9^{2(n+m)}/\delta_k)}{c\|\bar{\Phi}_{ck}^{-1}\|_\op^{-2}}.
\end{equation}
From Lemma \ref{lem:comparator_spectral_lines} and Proposition \ref{prop:FE_from_spectral_lines}, it follows that for each $k$, with probability at least $1 - \delta_k$,
\begin{equation} \label{eqn:comparator_info_bound}
    \sum_{t=t_k}^{t_{k+1}-1} \phi_{ct}\phi_{ct}^\top \geq \frac{\|\bar{\Phi}_{ck}^{-1}\|_\op^{-2}T_k}{2(n+m)}I_{n+m} \geq \frac{T_k}{2(n+m)\bar{C}_c^2(k+1)}I_{n+m}.
\end{equation}
Furthermore, we can rewrite $u_t$ in \eqref{eqn:control_input} as
\begin{equation*} \label{eqn:control_input_rewritten_2}
    u_t = v_t + \Theta_{B*}^{-1}(\Theta_{A*} + \Delta_{mk_t})e_{ct} - \Theta_{B*}^{-1}\tilde{\Theta}_t\phi_t,
\end{equation*}
and thus we have
\begin{equation*} \label{eqn:phi_vs_phi_c}
    \phi_t = \begin{bmatrix} -x_t \\ u_t \end{bmatrix} = \phi_{ct} + \underbrace{\begin{bmatrix} -I_n \\ \Theta_{B*}^{-1}(\Theta_{A*} + \Delta_{mk_t}) \end{bmatrix}}_{D_{ek}}e_{ct} + \underbrace{\begin{bmatrix} 0 \\ -\Theta_{B*}^{-1} \end{bmatrix}}_{D_\Theta}\tilde{\Theta}_t\phi_t
\end{equation*}
and
\begin{align}
    \sum_{t=t_k}^{t_{k+1}-1} \phi_t\phi_t^\top &= \sum_{t=t_k}^{t_{k+1}-1} \phi_{ct}\phi_{ct}^\top + \sum_{t=t_k}^{t_{k+1}-1} D_{ek}e_{ct}e_{ct}^\top D_{ek}^\top + \sum_{t=t_k}^{t_{k+1}-1} D_\Theta\tilde{\Theta}_t\phi_t\phi_t^\top\tilde{\Theta}_t^\top D_\Theta^\top \nonumber \\
    &\indenti{=} + \sum_{t=t_k}^{t_{k+1}-1} \phi_{ct}e_{ct}^\top D_{ek}^\top + \sum_{t=t_k}^{t_{k+1}-1} D_{ek}e_{ct}\phi_{ct}^\top + \sum_{t=t_k}^{t_{k+1}-1} \phi_{ct}\phi_t^\top\tilde{\Theta}_t^\top D_\Theta^\top \nonumber \\
    &\indenti{=} +\sum_{t=t_k}^{t_{k+1}-1}  D_\Theta\tilde{\Theta}_t\phi_t\phi_{ct}^\top + \sum_{t=t_k}^{t_{k+1}-1} D_{ek}e_{ct}\phi_t^\top\tilde{\Theta}_t^\top D_\Theta^\top + \sum_{t=t_k}^{t_{k+1}-1} D_\Theta\tilde{\Theta}_t\phi_te_{ct}^\top D_{ek}^\top \nonumber \\
    &\geq \frac{1}{2}\sum_{t=t_k}^{t_{k+1}-1} \phi_{ct}\phi_{ct}^\top - 4\sum_{t=t_k}^{t_{k+1}-1} D_{ek}e_{ct}e_{ct}^\top D_{ek}^\top - 4\sum_{t=t_k}^{t_{k+1}-1} D_\Theta\tilde{\Theta}_t\phi_t\phi_t^\top\tilde{\Theta}_t^\top D_\Theta^\top. \label{eqn:phi_phi^T_lower_bound}
\end{align}

Now, consider the sequence $\{\bar{T}_k\}_{k \geq 0}$ defined for each $k \geq 0$ as follows:
\begin{equation} \label{eqn:T_bar_k}
    \begin{gathered}
        \bar{T}_k = \inf_{\bar{T} \geq 0} \bar{T} \\
        \mathrm{s.t.} \\
        \bar{T} \geq \max\left\{\frac{2(n+m)\bar{\sigma}_c^2\ln(9^{2(n+m)}/\delta_k)\bar{C}_c^2(k+1)}{c}, C_T(k+1)\right\}, \\
        \frac{\bar{T}}{4(n+m)\bar{C}_c^2(k+1)} - 4\|D_{ek}\|_\op^2\sum_{t=t_k}^{t_k+\bar{T}-1} \|e_{ct}\|^2 - 4\|D_\Theta\|_\op^2\sum_{t=t_k}^{t_k+\bar{T}-1} \|\tilde{\Theta}_t\phi_t\|^2 \geq C_\Lambda.
    \end{gathered}
\end{equation}
We can show that $T_k \leq \mathrm{ceil}(\bar{T}_k)\ \forall k \geq 0$ with probability at least $1 - \delta'$ as follows. Fix any epoch $k$, consider the time step $T = \mathrm{ceil}(t_k + \bar{T}_k)$, and suppose that the time step $T - 1$ is contained in epoch $k$. Then, from \eqref{eqn:phi_phi^T_lower_bound}-\eqref{eqn:T_bar_k}, we see that $T \geq t_k + C_T(k+1)$, and $\sum_{t=t_k}^{T-1} \phi_t\phi_t^\top \geq C_\Lambda$ with probability at least $1 - \delta_k$. Thus, from Line \ref{algln:ref_model_update_conditions} of Algorithm \ref{alg:mrac-lqr}, we know that epoch $k$ will conclude, and we will have $T_k = \mathrm{ceil}(\bar{T}_k)$, with probability at least $1 - \delta_k$. The claim that $T_k \leq \mathrm{ceil}(\bar{T}_k)\ \forall k \geq 0$ with probability at least $1 - \delta'$ follows by taking a union bound over all $k \geq 0$ and noting that $\sum_{k=0}^\infty \delta_k = \delta'$ \citep{ayoub1974EulerZeta}.

Given that $T_k \leq \mathrm{ceil}(\bar{T}_k)\ \forall k \geq 0$ with probability at least $1 - \delta'$, all that remains in order to show the claim is to show that $\bar{T}_k \leq \tilde{O}(k+1)$. If this is true, then we also have $T_k \leq \tilde{O}(k+1)$, and it follows straightforwardly that $k_T \leq \tilde{O}(\sqrt{T})$.

Suppose that $\bar{T}_k = \tilde{O}((k+1)^{1+a})$ for any $a > 0$. Then, from \eqref{eqn:T_bar_k}, there will be an epoch $k_* \in \bbZ_{\geq 0}$ such that
\begin{equation*}
    \frac{\bar{T}_k}{4(n+m)\bar{C}_c^2(k+1)} - 4\|D_{ek}\|_\op^2\sum_{t=t_k}^{t_k+\bar{T}_k-1} \|e_{ct}\|^2 - 4\|D_\Theta\|_\op^2\sum_{t=t_k}^{t_k+\bar{T}_k-1} \|\tilde{\Theta}_t\phi_t\|^2 = C_\Lambda\ \forall k \geq k_*.
\end{equation*}
Summing over subsequent epochs and applying Propositions \ref{prop:confidence_set_convergence}-\ref{prop:regressor_stability} and Theorem \ref{thm:stability}, we obtain
\begin{align}
    \sum_{k=k_*}^{k_T-1} \left(\frac{\bar{T}_k}{4(n+m)\bar{C}_c^2(k+1)} - C_\Lambda\right) &= O\left(\sum_{k=k_*}^{k_T-1}\sum_{t=t_k}^{t_k+\bar{T}-1} \|e_{ct}\|^2 + \sum_{k=k_*}^{k_T-1}\sum_{t=t_k}^{t_k+\bar{T}-1} \|\tilde{\Theta}_t\phi_t\|^2\right) \nonumber \\
    &\leq O\left(\sum_{t=0}^{T-1} \|e_{ct}\|^2 + \sum_{t=0}^{T-1} \|\tilde{\Theta}_t\phi_t\|^2\right) \nonumber \\
    &\leq \tilde{O}(\sqrt{T}). \label{eqn:T_bar_k_growth_bound_1}
\end{align}
Additionally, from the definitions of $T_k$ and $k_T$, we can write
\begin{equation} \label{eqn:T_bar_k_growth_bound_2}
    T \leq \sum_{k=0}^{k_T} T_k \leq \sum_{k=0}^{k_T} \bar{T}_k = \sum_{k=0}^{k_T} \tilde{O}((k+1)^{1+a}) = \tilde{O}((k_T+1)^{2+a}).
\end{equation}
Combining \eqref{eqn:T_bar_k_growth_bound_1} and \eqref{eqn:T_bar_k_growth_bound_2}, we finally obtain
\begin{equation*}
    \sum_{k=k_*}^{\tilde{O}(T^{1/(2+a)})-1} \left(\tilde{O}((k+1)^a) - C_\Lambda\right) \leq \tilde{O}(\sqrt{T}),
\end{equation*}
which one may easily verify cannot hold for any $a > 0$. Thus, we reach a contradiction.
\qed}

\subsection{Guarantees of parameter convergence and uniform stability}

Our first bound on the parameter error is a bound within each epoch, which makes use of the reference model update conditions in Line \ref{algln:ref_model_update_conditions} of Algorithm \ref{alg:mrac-lqr}.
\begin{lemma} \label{lem:parameter_error_in_epoch_k}
    On the event $\calE_2$, with probability at least $1 - \delta'$ for any $\delta' \in (0, 1)$, we have $\|\tilde{\Theta}_t\|_F^2 \leq O(\frac{\mathrm{polylog}(1 + T)}{\lambda + k_tC_\Lambda})$ for all $t \in [0, T)$.
\end{lemma}
\proof{
Recall from Appendix \ref{app:confidence_set_convergence} that, on the event $\calE_2$, for any $\delta' \in (0, 1)$, \eqref{eqn:Xi_prime_bound} holds with $\Xi_t' = \hat{\Theta}_t$. Furthermore, we see from \eqref{eqn:RLS_cov_inv_update} that all eigenvalues of $\Sigma_t^{-1}$ are nondecreasing with $t$. Thus, considering any $t \in [0, T)$, we have $\Sigma_T^{-1} \succeq \Sigma_t^{-1} \succeq \lambda + k_tC_\Lambda$, where the last inequality comes from Line \ref{algln:ref_model_update_conditions} of Algorithm \ref{alg:mrac-lqr}. Therefore, with probability at least $1 - \delta'$ for any $\delta' \in (0, 1)$, we have
\begin{gather}
    (\lambda + k_tC_\Lambda)\|\tilde{\Theta}_t\|_F^2 \leq \Tr[\tilde{\Theta}_t\Sigma_t^{-1}\tilde{\Theta}_t^\top] \leq 4\beta_\delta(\Sigma_t) \leq 4\beta_\delta(\Sigma_T) \leq O\left(\ln\left(\frac{1 + C'T}{\delta^{2/d}}\right)\right) \label{eqn:theta_tilde_bound_2}
\end{gather}
where $C'$ is a constant derived from $\delta'$ (see Appendix \ref{app:confidence_set_convergence}).
\qed}

Note that the final inequality in \eqref{eqn:theta_tilde_bound_2} is very loose for small $t$, but it helps to simplify the analysis. Next, using our bound on the epoch lengths derived in the previous section, the following result guarantees that the parameter error eventually becomes small.




\begin{lemma} \label{lem:parameter_error_small_general}
    For any $\epsilon > 0$, on the event $\calE_2$, there exists a finite $T_\epsilon \in \bbZ_{\geq 0}$ with probability one such that $\|\tilde{\Theta}_t\|_\op \leq \epsilon$ for all $t \geq T_\epsilon$.
\end{lemma}
\begin{proof}
    Suppose not. Then, on the event $\calE_2$, $\exists \epsilon > 0$ such that, with nonzero probability, given any $T_\epsilon \in \bbZ_{\geq 0}$, $\exists t \geq T_\epsilon$ with $\|\tilde{\Theta}_t\|_\op > \epsilon$. Recall from Appendix \ref{app:confidence_set_convergence} that, on the event $\calE_2$, for any $\delta' \in (0, 1)$, \eqref{eqn:Xi_prime_bound} holds with $\Xi_t' = \hat{\Theta}_t$. If Lemma \ref{lem:parameter_error_small_general} is not true, therefore, \eqref{eqn:Xi_prime_bound} implies that $\Sigma_t^{-1}$ must have an eigenvalue which increases at most logarithmically with $t$ with nonzero probability. It follows from Line \ref{algln:ref_model_update_conditions} of Algorithm \ref{alg:mrac-lqr} that $T_k$ must increase at least exponentially with $k$ with nonzero probability, contradicting Lemma \ref{lem:epochs_linear}.
\end{proof}

Define $\epsilon_* = \min\{\frac{1}{16\|B_m\|_\op\|P_*\|_\op^2}, \frac{1}{40(2^{7/4})\|B_m\|_\op\|P_*\|_\op^5 + 10\|P_*\|_\op^{3/2}\sup_{t \geq 0} \|B_*\hat{\Theta}_{Bt}^{-1}\|_\op(1 + \|\hat{K}_{k_t}\|_\op)}\}$, and note that Lemma \ref{lem:gains_bounded} guarantees that $\epsilon_* > 0$. Further define $T_{\rm learn} = \inf_{k \geq 0} t_k$ such that $T_{\rm learn} \geq T_{\epsilon_*}$, and $k_{\rm learn} = k_{T_{\rm learn}}$. The following result bounds the error between the time-varying gain $\bar{K}_t$ in \eqref{eqn:control_input_rewritten} and the optimal gain.
\begin{lemma} \label{lem:K_bar_bound}
    On the event $\calE_2$, for all $k \geq k_{\rm learn}$ and all $t \in [t_k, t_{k+1})$, the following hold for norms $\circ \in \{\op, F\}$:
    \begin{enumerate}
        \item $\|\sqrt{R}(\bar{K}_t - K_*)\|_\circ \leq \left(7\|B_m\|_\circ(2\|P_*\|_\op^2)^{7/4} + \|\sqrt{R}\hat{\Theta}_{Bt}^{-1}\|_\circ(\|I_n\|_\circ + \|\hat{K}_{k_t}\|_\circ)\right)\|\tilde{\Theta}_{t_{k_t}}\|_\circ \\
        \indenti{\|\sqrt{R}(\bar{K}_t - K_*)\|_\circ \leq} + \|\sqrt{R}\hat{\Theta}_{Bt}^{-1}\|_\circ(\|I_n\|_\circ + \|\hat{K}_{k_t}\|_\circ)\|\tilde{\Theta}_t\|_\circ$, and
        \item $\|B_*(\bar{K}_t - K_*)\|_\circ \leq \left(8\|B_m\|_\circ(2\|P_*\|_\op^2)^{7/4} + \|B_*\hat{\Theta}_{Bt}^{-1}\|_\circ(\|I_n\|_\circ + \|\hat{K}_{k_t}\|_\circ)\right)\|\tilde{\Theta}_{t_{k_t}}\|_\circ \\
        \indenti{\|B_*(\bar{K}_t - K_*)\|_\circ \leq} + \|B_*\hat{\Theta}_{Bt}^{-1}\|_\circ(\|I_n\|_\circ + \|\hat{K}_{k_t}\|_\circ)\|\tilde{\Theta}_t\|_\circ$.
    \end{enumerate}
\end{lemma}
\proof{
Fix any $k \geq k_{\rm learn}$. Then, we have $t \geq T_{\rm learn}$ and thus $\|\tilde{\Theta}_t\|_\op \leq \frac{1}{16\|B_m\|_\op\|P_*\|_\op^2}$ for all $t \in [t_k, t_{k+1})$. Therefore, noting that $\hat{A}_k - A_* = -B_m\tilde{\Theta}_{At_k}$ and $\hat{B}_k - B_* = B_m\tilde{\Theta}_{Bt_k}$ from Assumption \ref{asn:matching_condition} and Algorithm \ref{alg:mrac-lqr}, we have
\begin{equation*}
    \min\{\|\hat{A}_k - A_*\|_\op, \|\hat{B}_k - B_*\|_\op\} \leq \|B_m\tilde{\Theta}_{t_k}\|_\op \leq \frac{1}{16\|P_*\|_\op^2}.
\end{equation*}
The claim follows from Proposition \ref{prop:dynamics_error_to_gain_error}, \eqref{eqn:control_input_rewritten}, and straightforward algebra.
\qed}




\subsection{Proof of Proposition \ref{prop:conditions_and_stability}} \label{app:conditions_and_stability}

For part (i) of Condition \ref{cond:TVRM}, as $\Delta_{m0} = 0$, we need only show that $A_{mk} = A_m + B_m\Delta_{mk} \forall k \geq 1$. From lines \ref{algln:dynamics_est}-\ref{algln:gain_offset} of Algorithm \ref{alg:mrac-lqr}, we have
\begin{equation*}
    A_{mk} = \hat{A}_k + \hat{B}_k\hat{K}_k = A_m - B_m\hat{\Theta}_{At_k} + B_m\hat{\Theta}_{Bt_k}\hat{K}_k = A_m + B_m\Delta_{mk}.
\end{equation*}
Parts (ii) and (iii) are proven in Lemma \ref{lem:gains_bounded}, and part (iv) is guaranteed by Line \ref{algln:ref_model_update_conditions} of Algorithm \ref{alg:mrac-lqr}.

For Condition \ref{cond:theta_in_set}, from Line \ref{algln:parameter_set} of Algorithm \ref{alg:mrac-lqr}, we have $Z_t = S_\Theta \cap C_\delta(\Xi_t, \Sigma_t)\ \forall t \geq 0$. By Assumption \ref{asn:params_bounded}, we have $\Theta_* \in S_\Theta$. Therefore, the event $\calE_2$ holds if and only if $\calE_1$ in Proposition \ref{prop:confidence_sets} holds, which occurs with probability at least $1 - \delta$.

Finally, the claim $\lim_{T \to \infty} \frac{1}{T}\sum_{t=1}^T \|e_{ct}\|^2 = 0$ follows directly from Proposition \ref{prop:confidence_set_convergence}, Proposition \ref{prop:regressor_stability}, and Theorem \ref{thm:stability}.
\qed

\subsection{Proof of Theorem \ref{thm:regret}} \label{app:regret}

Throughout this proof, we will suppose that the event $\calE_2$ as in Proposition \ref{prop:conditions_and_stability} holds. Combining \eqref{eqn:plant} and \eqref{eqn:control_input_rewritten}, the closed-loop dynamics can be written as
\begin{equation}
    x_{t+1} = (A_* + B_*\bar{K}_t)x_t + B_*\hat{\Theta}_{Bt}^{-1}r_t + w_{t+1}.
\end{equation}
Leveraging the superposition principle, we consider two signals $x_{wt}$ and $x_{rt}$ such that
\begin{align}
    x_t &= x_{wt} + x_{rt}, \label{eqn:x_decomposition} \\
    x_{w(t+1)} &= (A_* + B_*\bar{K}_t)x_{wt} + w_{t+1}, \label{eqn:x_w_dynamics} \\
    x_{r(t+1)} &= (A_* + B_*\bar{K}_t)x_{rt} + B_*\hat{\Theta}_{Bt}^{-1}r_t. \label{eqn:x_r_dynamics}
\end{align}
Now, suppose that $T > T_{\rm learn}$. Using \eqref{eqn:control_input_rewritten} and \eqref{eqn:x_decomposition} and straightforward algebra, we propose the following decomposition of the regret in \eqref{eqn:regret_definition}:
\begin{equation} \label{eqn:regret_decomposition}
    \begin{aligned}
        \Regret(T) = &\underbrace{\sum_{t=0}^{T_{\rm learn}-1} \left(x_t^\top Qx_t + u_t^\top Ru_t - J_*\right)}_{R_0} + \underbrace{\sum_{t=T_{\rm learn}}^{T-1} \left(x_{wt}^\top(Q + \bar{K}_t^\top R\bar{K}_t)x_{wt} - J_*\right)}_{R_1(T)} \\
        &+ \underbrace{\sum_{t=T_{\rm learn}}^{T-1} \left(x_{rt}^\top(Q + \bar{K}_t^\top R\bar{K}_t)x_{rt} + 2x_{rt}^\top\bar{K}_t^\top R\hat{\Theta}_{Bt}^{-1}r_t + r_t^\top\hat{\Theta}_{Bt}^{-\top}R\hat{\Theta}_{Bt}^{-1}r_t\right)}_{R_2(T)} \\
        &+ \underbrace{\sum_{t=T_{\rm learn}}^{T-1} 2x_{wt}^\top\left((Q + \bar{K}_t^\top R\bar{K}_t)x_{rt} + \bar{K}_t^\top R\hat{\Theta}_{Bt}^{-1}r_t\right)}_{R_3(T)}.
    \end{aligned}
\end{equation}
We see straightforwardly from Theorems \ref{thm:TVRM_stability} that $R_0 = O(1)$ with probability one.
What remains is to bound $R_1(T)$, $R_2(T)$, and $R_3(T)$.

\subsubsection*{Bounding $R_1(T)$}

For ease of notation, denote $\bar{P}_t = P_{\infty(A_*, B_*)}[\bar{K}_t]$, where $P_{\infty(A_*, B_*)}[\bar{K}_t]$ is defined in \eqref{eqn:dlyap_cost}.
Define the sequence $Z_t = x_{wt}^\top\bar{P}_tx_{wt}\ \forall t \geq T_{\rm learn}$. Using \eqref{eqn:dlyap_cost} and \eqref{eqn:x_w_dynamics}, the increment of $Z_t$ is given by
\begin{equation} \label{eqn:Z_t_increment}
    \begin{aligned}
        Z_{t+1} - Z_t &= -x_{wt}^\top(Q + \bar{K}_t^\top R\bar{K}_t)x_{wt} + 2x_{wt}^\top(A_* + B_*\bar{K}_t)^\top\bar{P}_tw_{t+1} + w_{t+1}^\top\bar{P}_tw_{t+1} \\
        &\indenti{=} + x_{w(t+1)}^\top(\bar{P}_{t+1} - \bar{P}_t)x_{w(t+1)}.
    \end{aligned}
\end{equation}
Using \eqref{eqn:Z_t_increment}, we can then rewrite $R_1(T)$ as
\begin{equation} \label{eqn:R_1_rewritten}
    \begin{aligned}
        R_1(T) &= Z_T - Z_{T_{\rm learn}} + \sum_{t=T_{\rm learn}}^{T-1} \left(2x_{wt}^\top(A_* + B_*\bar{K}_t)^\top\bar{P}_tw_{t+1} + w_{t+1}^\top P_*w_{t+1} - J_*\right) \\
        &\indenti{=} + \sum_{t=T_{\rm learn}}^{T-1} \left(w_{t+1}^\top(\bar{P}_t - P_*)w_{t+1} + x_{w(t+1)}^\top(\bar{P}_{t+1} - \bar{P}_t)x_{w(t+1)}\right).
    \end{aligned}
\end{equation}
For all $t \geq T_{\rm learn}$, combining Lemma \ref{lem:K_bar_bound}(ii) with the definition of $\epsilon_*$, we have $\|B_*(\bar{K}_t - K_*)\|_\op \leq \frac{1}{5\|P_*\|_\op^{3/2}}$, and thus Proposition \ref{prop:optimality_of_gain}(i) holds with $K = \bar{K}_t$. From Proposition \ref{prop:optimality_of_gain}(i) and Lemma \ref{lem:gains_bounded}, $\bar{P}_t$ is uniformly bounded for all $t \geq T_{\rm learn}$. Additionally, as in the proof of Proposition \ref{prop:regressor_stability}, $\|x_{wt}\|^2$ is a sub-exponential random variable with uniformly bounded norm. Therefore, for any $\delta' \in (0, 1)$, $Z_T \leq O(\ln(\frac{2}{\delta'}))$ with probability at least $1 - \delta'$. Bounding $-Z_{T_{\rm learn}}$ is trivial, as $\bar{P}_t > 0$ and therefore $-Z_{T_{\rm learn}} \leq 0$.

Define $\xi_t = 2x_{wt}^\top(A_* + B_*\bar{K}_t)^\top\bar{P}_tw_{t+1} + w_{t+1}^\top P_*w_{t+1} - J_*$. Recall that $\bbE[w_{t+1} | \calF_t] = 0$ and that $x_{wt}$ and $\bar{K}_t$ are $\calF_t$-measurable. Additionally, it is well-known \citep[see e.g.,][]{stengel1994optimal} that $J_* = \Tr[P_*\Sigma_w] = \bbE[w_{t+1}^\top P_*w_{t+1} | \calF_t]$. Therefore, $\bbE[\xi_t | \calF_t] = 0$, and $\xi_t$ is a martingale difference sequence. We thus immediately apply Proposition \ref{prop:mds_bound} to obtain
\begin{equation} \label{eqn:R_1_first_sum_bound}
    \sum_{t=T_{\rm learn}}^{T-1} \xi_t = \sum_{t=T_{\rm learn}}^{T-1} \left(2x_{wt}^\top(A_* + B_*\bar{K}_t)^\top\bar{P}_tw_{t+1} + w_{t+1}^\top P_*w_{t+1} - J_*\right) = \tilde{O}(\sqrt{T})
\end{equation}
with probability one.

To bound the final sum in \eqref{eqn:R_1_rewritten}, we first note that $\|w_{t+1}\|^2$ and $\|x_{w(t+1)}\|^2$ are sub-exponential random variables with uniformly bounded norms, and therefore that, using Lemma \ref{lem:subE_tail_bound} and appropriate union bounds, for any $\delta'' \in (0, 1)$, we have $\max_{t \in [T_{\rm learn}, T)} \|w_{t+1}\|^2 \leq O(\ln(\frac{4(T - T_{\rm learn})}{\delta''}))$ and $\max_{t \in [T_{\rm learn}, T)} \|x_{w(t+1)}\|^2 \leq O(\ln(\frac{4(T - T_{\rm learn})}{\delta''}))$ with probability at least $1 - \delta''$. Then, using Proposition \ref{prop:optimality_of_gain}(i), Lemma \ref{lem:K_bar_bound} with Lemma \ref{lem:gains_bounded}, and Lemma \ref{lem:parameter_error_in_epoch_k}, we have
\begin{equation} \label{eqn:P_bar_bound}
    \|\bar{P}_t - P_*\|_\op \leq O(\|\tilde{\Theta}_{t_{k_t}}\|_\op + \|\tilde{\Theta}_t\|_\op) \leq O\left(\frac{\mathrm{polylog}(1 + T)}{\lambda + k_tC_\Lambda}\right)
\end{equation}
with probability at least $1 - \delta'''$ for any $\delta''' \in (0, 1)$.

Finally, putting \eqref{eqn:R_1_rewritten} together with \eqref{eqn:R_1_first_sum_bound}, \eqref{eqn:P_bar_bound}, and our high-probability bounds on $Z_T$, $\max_{t \in [T_{\rm learn}, T)} \|w_{t+1}\|^2$, and $\max_{t \in [T_{\rm learn}, T)} \|x_{w(t+1)}\|^2$, we obtain, with probability at least $1 - \delta' - \delta'' - \delta'''$,
\begin{align}
    R_1(T) &\leq O\left(\ln\left(\frac{2}{\delta'}\right)\right) + \tilde{O}(\sqrt{T}) + \sum_{t=T_{\rm learn}}^{T-1} O\left(\ln\left(\frac{4(T - T_{\rm learn})}{\delta''}\right)\frac{\mathrm{polylog}(1 + T)}{\lambda + k_tC_\Lambda}\right) \nonumber \\
    &\leq O(1) + \tilde{O}(\sqrt{T}) + \sum_{k=k_{\rm learn}}^{k_T} O(\mathrm{polylog}(1 + T)T_k(k + 1)^{-1}) \label{eqn:R_1_bound}
\end{align}

\subsubsection*{Bounding $R_2(T)$}

Lemma \ref{lem:gains_bounded} guarantees that $\|\bar{K}_t\|_\op$ and $\|\hat{\Theta}_{Bt}^{-1}\|_\op$ are uniformly bounded, allowing us to write
\begin{equation} \label{eqn:R_2_bound_1}
    R_2(T) \leq \sum_{t=T_{\rm learn}}^{T-1} O(\|x_{rt}\|^2 + \|x_{rt}\|\|r_t\| + \|r_t\|^2).
\end{equation}
From Line \ref{algln:r_t} of Algorithm \ref{alg:mrac-lqr}, we have $\|r_t\| = O((k_t+1)^{-1/2})\ \forall t \geq 0$. Furthermore, Proposition \ref{prop:optimality_of_gain}(ii) allows us to show that $\|x_{rt}\|$ decays at the same rate as $\|r_t\|$, as follows. For all $t \geq T_{\rm learn}$, combining Lemma \ref{lem:K_bar_bound}(ii) with the definition of $\epsilon_*$, we have $\|B_*(\bar{K}_t - K_*)\|_\op \leq \frac{1}{5\|P_*\|_\op^{3/2}}$, and thus Proposition \ref{prop:optimality_of_gain}(ii) holds with $K = \bar{K}_t$. Define $X_t = x_{rt}^\top P_{\rm lyap}x_{rt}$. Then, using \eqref{eqn:x_r_dynamics}, for any $k \geq k_{\rm learn}$ and $t \in [t_k, t_{k+1})$, we have
\begin{align}
    X_{t+1} &= X_t + x_{rt}^\top((A_* + B_*\bar{K}_t)^\top P_{\rm lyap}(A_* + B_*\bar{K}_t) - P_{\rm lyap})x_{rt} \nonumber \\
    &\indenti{= X_t} + 2x_{rt}^\top(A_* + B_*\bar{K}_t)^\top P_{\rm lyap}B_*\hat{\Theta}_{Bt}^{-1}r_t + r_t^\top\hat{\Theta}_{Bt}^{-\top}B_*^\top P_{\rm lyap}B_*\hat{\Theta}_{Bt}^{-1}r_t \nonumber \\
    &\leq \left(1 - \frac{1}{2\|P_{\rm lyap}\|_\op}\right)X_t + 2d_1\|\hat{\Theta}_{Bt}^{-1}r_t\|\sqrt{X_t} + d_2\|\hat{\Theta}_{Bt}^{-1}r_t\|^2 \nonumber \\
    &\leq d_3X_t + d_4\|\hat{\Theta}_{Bt}^{-1}r_t\|^2 \label{eqn:X_difference_equation}
\end{align}
where
\begin{align*}
    d_1 &= \sup_{t \geq 0} \|\sqrt{P_{\rm lyap}}^{-1}(A_* + B_*\bar{K}_t)^\top P_{\rm lyap}B_*\|_\op, \\
    d_2 &= \|B_*^\top P_{\rm lyap}B_*\|_\op, \\
    d_3 &= 1 - \frac{1}{4\|P_{\rm lyap}\|_\op}, \\
    d_4 &= 4d_1^2\|P_{\rm lyap}\|_\op + d_2
\end{align*}
and \eqref{eqn:X_difference_equation} uses the relation $-ax + 2b\sqrt{x} + c \leq -\frac{a}{2}x + \frac{2b^2}{a} + c$. Then,
\begin{align}
    X_t &\leq d_3^{t - t_k}X_{t_k} + d_4\sum_{s=t_k}^{t-1} d_3^{t-1-s}\|\hat{\Theta}_{Bt}^{-1}r_t\|^2 \nonumber \\
    &\leq d_3^{t - t_k}X_{t_k} + \frac{d_4}{1 - d_3}\max_{t \in [t_k, t_{k+1})}\|\hat{\Theta}_{Bt}^{-1}r_t\|^2\ \forall t \in (t_k, t_{k+1}] \implies \nonumber \\
    \|x_{rt}\| &\leq O(d_3^{(t - t_k)/2}\|x_{rt_k}\|) + O\left(\max_{t \in [t_k, t_{k+1})}\|\hat{\Theta}_{Bt}^{-1}r_t\|\right)\ \forall t \in (t_k, t_{k+1}]. \label{eqn:x_rt_rate_of_decay}
\end{align}
Finally, from \eqref{eqn:R_2_bound_1}, \eqref{eqn:x_rt_rate_of_decay}, the fact that $d_3 \in (0, 1)$, the fact that $\|x_{rt_k}\|^2 + \|x_{rt_k}\| \leq O(1)$, and the fact that $\max_{t \in [t_k, t_{k+1})}\|\hat{\Theta}_{Bt}^{-1}r_t\| = O((k+1)^{-1/2})$, we have
\begin{equation} \label{eqn:R_2_bound}
    R_2(T) \leq \sum_{k=k_{\rm learn}}^{k_T} \left(O(1) + O(T_k(k+1)^{-1})\right).
\end{equation}

\subsubsection*{Bounding $R_3(T)$}

We must take care in bounding $R_3(T)$, as $x_{wt}$ and $\bar{K}_t$ are both random and not independent. Define $\Delta K_t := \bar{K}_t - \hat{K}_{k_t} = \hat{\Theta}_{Bt}^{-1}((\tilde{\Theta}_{Bt_{k_t}} - \tilde{\Theta}_{Bt})\hat{K}_{k_t} + \tilde{\Theta}_{At} - \tilde{\Theta}_{At_{k_t}})$, and $\Delta\hat{\Theta}_{Bt}^{-1} := \hat{\Theta}_{Bt}^{-1} - \hat{\Theta}_{Bt_{k_t}}^{-1} = \hat{\Theta}_{Bt_{k_t}}^{-1}(\tilde{\Theta}_{Bt_{k_t}} - \tilde{\Theta}_{Bt})\hat{\Theta}_{Bt}^{-1}$. Then, leveraging the superposition principle, we consider four signals - $x_{wt}^{(1)}$, $w_{wt}^{(2)}$, $x_{rt}^{(1)}$, and $x_{rt}^{(2)}$ - such that
\begin{align}
    x_{wt} &= x_{wt}^{(1)} + x_{wt}^{(2)}, \label{eqn:x_w_decomposition} \\
    x_{rt} &= x_{rt}^{(1)} + x_{rt}^{(2)}, \label{eqn:x_r_decomposition} \\
    x_{w(t+1)}^{(1)} &= (A_* + B_*\hat{K}_{k_t})x_{wt}^{(1)} + w_{t+1}, \label{eqn:x_w_1_dynamics} \\
    x_{w(t+1)}^{(2)} &= (A_* + B_*\hat{K}_{k_t})x_{wt}^{(2)} + B_*\Delta K_tx_{wt}, \label{eqn:x_w_2_dynamics} \\
    x_{r(t+1)}^{(1)} &= (A_* + B_*\hat{K}_{k_t})x_{rt}^{(1)} + B_*\hat{\Theta}_{Bt_{k_t}}^{-1}r_t, \label{eqn:x_r_1_dynamics} \\
    x_{r(t+1)}^{(2)} &= (A_* + B_*\hat{K}_{k_t})x_{rt}^{(2)} + B_*\Delta\hat{\Theta}_{Bt}^{-1}r_t + B_*\Delta K_tx_{rt}. \label{eqn:x_r_2_dynamics}
\end{align}
Then, $R_3(T)$ can be further decomposed as
\begin{equation} \label{eqn:R_3_decomposition}
    \begin{aligned}
        R_3(T) = &\underbrace{\sum_{t=T_{\rm learn}}^{T-1} 2(x_{wt}^{(1)})^\top\left((Q + \hat{K}_{k_t}^\top R\hat{K}_{k_t})x_{rt}^{(1)} + \hat{K}_{k_t}^\top R\hat{\Theta}_{Bt_{k_t}}^{-1}r_t\right)}_{R_{3,1}(T)} \\
        &\underbrace{+ \sum_{t=T_{\rm learn}}^{T-1} \begin{aligned}
            2(x_{wt}^{(1)})^\top\Big(&(Q + \hat{K}_{k_t}^\top R\hat{K}_{k_t})x_{rt}^{(2)} \\
            &+ (\Delta K_t^\top R\hat{K}_{k_t} + \hat{K}_{k_t}^\top R\Delta K_t + \Delta K_t^\top R\Delta K_t)x_{rt} \\
            &+ (\Delta K_t^\top R\hat{\Theta}_{Bt_{k_t}}^{-1} + \hat{K}_{k_t}^\top R\Delta\hat{\Theta}_{Bt}^{-1} + \Delta K_t^\top R\Delta\hat{\Theta}_{Bt}^{-1})r_t\Big)
        \end{aligned}}_{R_{3,2}(T)} \\
        &+ \underbrace{\sum_{t=T_{\rm learn}}^{T-1} 2(x_{wt}^{(2)})^\top\left((Q + \bar{K}_t^\top R\bar{K}_t)x_{rt} + \bar{K}_t^\top R\hat{\Theta}_{Bt}^{-1}r_t\right)}_{R_{3,3}(T)}.
    \end{aligned}
\end{equation}

\noindent\textbf{Bounding $\mathbf{R_{3,1}(T)}$} \\

\noindent For ease of notation, denote $A_{cl,k} = A_* + B_*\hat{K}_k$, and define $b_t = 2(Q + \hat{K}_{k_t}^\top R\hat{K}_{k_t})x_{rt}^{(1)} + \hat{K}_{k_t}^\top R\hat{\Theta}_{Bt_{k_t}}^{-1}r_t$. For any $k \geq k_{\rm learn}$ and any $t \in (t_k, t_{k+1}]$, using \eqref{eqn:x_w_1_dynamics}, we can write
\begin{equation*}
    x_{wt}^{(1)} = A_{cl,k}^{t-t_k}x_{wt_k}^{(1)} + \sum_{s=t_k}^{t-1} A_{cl,k}^{t-s-1}w_{s+1}
\end{equation*}
and thus, defining $\tau_k = \min\{T, t_{k+1}\}$, $m_{0,k} = \sum_{t=t_k}^{\tau_k-1} b_t^\top A_{cl,k}^{t-t_k}$, and $m_{t,k} = \sum_{s=t}^{\tau_k-1} b_s^\top A_{cl,k}^{s-t} \\
\forall t \in [t_k+1, \tau_k-1]$, we have
\begin{align*}
    R_{3,1}(T) = \sum_{t=T_{\rm learn}}^{T-1} b_t^\top x_{wt}^{(1)} &= \sum_{k=k_{\rm learn}}^{k_T} \sum_{t=t_k}^{\tau_k-1} b_t^\top\left(A_{cl,k}^{t-t_k}x_{wt_k}^{(1)} + \sum_{s=t_k}^{t-1} A_{cl,k}^{t-s-1}w_{s+1}\right) \\
    &= \sum_{k=k_{\rm learn}}^{k_T} \left(m_{0,k}x_{wt_k}^{(1)} + \sum_{t=t_k+1}^{\tau_k-1} m_{t,k}w_t\right).
    %
\end{align*}
Now, note that, for all $t \geq T_{\rm learn}$, combining Proposition \ref{prop:dynamics_error_to_gain_error}(ii) with the definition of $\epsilon_*$, we have $\|B_*(\hat{K}_{k_t} - K_*)\|_\op \leq \frac{1}{5\|P_*\|_\op^{3/2}}$, and thus Proposition \ref{prop:optimality_of_gain}(ii) holds with $K = \hat{K}_{k_t}$. It follows that $A_{cl,k}$ is Schur-stable for $k \geq k_{\rm learn}$, and thus that $\exists \gamma_{cl,k} \geq 1, \lambda_{cl,k} \in (0, 1)$ such that $\|A_{cl,k}^i\|_\op \leq \gamma_{cl,k}\lambda_{cl,k}^i\ \forall i \geq 0$. Additionally, \eqref{eqn:x_rt_rate_of_decay} holds for $\|x_{rt}^{(1)}\|$, replacing $\|x_{rt_k}\|$ with $\|x_{rt_k}^{(1)}\|$ and $\|\hat{\Theta}_{Bt}^{-1}r_t\|$ with $\|\hat{\Theta}_{Bt_{k_t}}^{-1}r_t\|$. Therefore, applying Lemma \ref{lem:gains_bounded}, we have
\begin{align}
    \|m_{0,k}\| &\leq \sum_{t=t_k}^{\tau_k-1} \gamma_{cl,k}\lambda_{cl,k}^{t-t_k}O\left(d_3^{(t-t_k)/2}\|x_{rt_k}^{(1)}\| + \max_{t \in [t_k, \tau_k)}\|\hat{\Theta}_{Bt_{k_t}}^{-1}r_t\|\right) \nonumber \\
    &\leq O\left(\|x_{rt_k}^{(1)}\| + (k+1)^{-1/2}\right), \label{eqn:m_0k_bound} \\
    \|m_{t,k}\| &\leq \sum_{s=t}^{\tau_k-1} \gamma_{cl,k}\lambda_{cl,k}^{s-t}O\left(d_3^{(s-t_k)/2}\|x_{rt_k}^{(1)}\| + \max_{t \in [t_k, \tau_k)}\|\hat{\Theta}_{Bt_{k_t}}^{-1}r_t\|\right) \nonumber \\
    &\leq O\left(d_3^{(t-t_k)/2}\|x_{rt_k}^{(1)}\| + (k+1)^{-1/2}\right) \implies \nonumber \\
    \sum_{t=t_k+1}^{\tau_k-1} \|m_{t,k}\|^2 &\leq \sum_{t=t_k+1}^{\tau_k-1} O\left(d_3^{t-t_k}\|x_{rt_k}^{(1)}\| + (k+1)^{-1}\right) \leq O\left(\|x_{rt_k}^{(1)}\| + T_k(k+1)^{-1}\right). \label{eqn:sum_m_tk_bound}
\end{align}
Finally, we observe that $w_t$ are independent zero-mean sub-Gaussian random vectors, and that each $w_t, t \in [t_k+1, \tau_k-1]$ is independent of each $m_{t,k}, t \in [t_k+1, \tau_k-1]$. Therefore, choosing any $\delta' \in (0, 1)$ and defining $\delta_k = \frac{6\delta'}{\pi^2(k+1)^2}$ as in the proof of Lemma \ref{lem:epochs_linear}, we can use Proposition \ref{prop:Hoeffding} with \eqref{eqn:sum_m_tk_bound} to bound
\begin{align}
    \sum_{t=t_k+1}^{\tau_k-1} m_{t,k}w_t &\leq \sqrt{\frac{\sigma_w^2\sum_{t=t_k+1}^{\tau_k-1} \|m_{t,k}\|^2}{C_2''}\ln\left(\frac{2}{\delta_k}\right)} \nonumber \\
    &\leq \sqrt{\frac{\sigma_w^2O\left(\|x_{rt_k}^{(1)}\| + T_k(k+1)^{-1}\right)}{C_2''}\ln\left(\frac{2}{\delta_k}\right)}. \label{eqn:sum_m_w_bound}
\end{align}
$\|x_{wt_k}^{(1)}\|$ is a sub-Gaussian random variable with uniformly bounded norm, and thus, applying Lemma \ref{lem:subG_tail_bound} with a union bound, for any $\delta'' \in (0, 1)$, we can bound $\max_{t \in [T_{\rm learn}, T)} \|x_{wt_k}^{(1)}\| \leq O(\ln(\frac{2T}{\delta''}))$. Additionally, by a union bound and the fact that $\sum_{k=0}^\infty \delta_k = \delta'$, \eqref{eqn:sum_m_w_bound} holds for all $k \geq k_{\rm learn}$ with probability at least $1 - \delta'$. We can thus use \eqref{eqn:m_0k_bound} and \eqref{eqn:sum_m_w_bound} to obtain, with probability at least $1 - \delta' - \delta''$,
\begin{equation} \label{eqn:R_3_1_bound}
    R_{3,1}(T) \leq \sum_{k=k_{\rm learn}}^{k_T} O\left(\ln\left(\frac{2T}{\delta''}\right)\left(1 + (k+1)^{-1/2}\right) + \sqrt{\tilde{O}\left(1 + T_k(k+1)^{-1}\right)}\right).
\end{equation}

\noindent\textbf{Bounding $\mathbf{R_{3,2}(T)}$} \\

\noindent As Proposition \ref{prop:optimality_of_gain}(ii) holds with $K = \hat{K}_{k_t}$, \eqref{eqn:x_rt_rate_of_decay} holds for $\|x_{rt}^{(2)}\|$, replacing $\|x_{rt_k}\|$ with $\|x_{rt_k}^{(2)}\|$ and $\|\hat{\Theta}_{Bt}^{-1}r_t\|$ with $\|\Delta\hat{\Theta}_{Bt}^{-1}r_t + \Delta K_tx_{rt}\|$. Leveraging this relation and Lemma \ref{lem:gains_bounded}, we can bound $R_{3,2}(T)$ as
\begin{align}
    R_{3,2}(T) &\leq \sum_{t=T_{\rm learn}}^{T-1} O\left(\|x_{wt}^{(1)}\|\left(\|x_{rt}^{(2)}\| + \|\Delta K_t\|_\op\|x_{rt}\| + \|\Delta K_t\|_\op\|r_t\| + \|\Delta\hat{\Theta}_{Bt}^{-1}\|_\op\|r_t\|\right)\right) \nonumber \\
    &\begin{aligned}
        \leq \sum_{k=k_{\rm learn}}^{k_T}\sum_{t=t_k}^{t_{k+1}-1} O\bigg(\|x_{wt}^{(1)}\|\bigg(&d_3^{(t - t_k)/2}\|x_{rt_k}^{(2)}\| + \max_{t \in [t_k, t_{k+1})}\|\Delta\hat{\Theta}_{Bt}^{-1}r_t + \Delta K_tx_{rt}\| \\
        &+ \|\Delta K_t\|_\op\|x_{rt}\| + \|\Delta K_t\|_\op\|r_t\| + \|\Delta\hat{\Theta}_{Bt}^{-1}\|_\op\|r_t\|\bigg)\bigg)
    \end{aligned} \nonumber \\
    &\leq \left(\max_{t \in [T_{\rm learn}, T)}\|x_{wt}^{(1)}\|\right) \nonumber \\
    &\indenti{\leq} \times \sum_{k=k_{\rm learn}}^{k_T} O\left(\|x_{rt_k}^{(2)}\| + T_k\max_{t \in [t_k, t_{k+1})}\left(\|\Delta K_t\|_\op(\|x_{rt}\| + \|r_t\|) + \|\Delta\hat{\Theta}_{Bt}^{-1}\|_\op\|r_t\|\right)\right). \label{eqn:R_3_2_bound_1}
\end{align}
Applying Lemma \ref{lem:gains_bounded}, we see that $\|\Delta K_t\|_\op = O(\|\tilde{\Theta}_{t_{k_t}}\|_\op + \|\tilde{\Theta}_t\|_\op)$ and $\|\Delta\hat{\Theta}_{Bt}^{-1}\|_\op = O(\|\tilde{\Theta}_{t_{k_t}}\|_\op + \|\tilde{\Theta}_t\|_\op)$. Then, using Lemma \ref{lem:parameter_error_in_epoch_k}, we obtain
\begin{align}
    \|\Delta K_t\|_\op &= O\left(\mathrm{polylog}(1 + T)(k_t + 1)^{-1/2}\right), \label{eqn:Delta_K_bound} \\
    \|\Delta\hat{\Theta}_{Bt}^{-1}\|_\op &= O\left(\mathrm{polylog}(1 + T)(k_t + 1)^{-1/2}\right). \label{eqn:Delta_Theta_B_bound}
\end{align}
Finally, it is straightforward to see that $\|x_{wt}^{(1)}\|$ is a sub-Gaussian random variable with uniformly bounded norm for $t \in [T_{\rm learn}, T)$, and thus, with probability at least $1 - \delta'$ for any $\delta' \in (0, 1)$, we have $\max_{t \in [T_{\rm learn}, T)} \|x_{wt}^{(1)}\| \leq O(\ln(\frac{2T}{\delta'}))$. Combining \eqref{eqn:R_3_2_bound_1}-\eqref{eqn:Delta_Theta_B_bound} with \eqref{eqn:x_rt_rate_of_decay} and the fact that $\|r_t\| = O((k_t + 1)^{-1/2})$, we obtain
\begin{equation} \label{eqn:R_3_2_bound}
    R_{3,2}(T) \leq O\left(\mathrm{polylog}(1 + T)\sum_{k=k_{\rm learn}}^{k_T} \left(1 + T_k(k+1)^{-1}\right)\right).
\end{equation}

\noindent\textbf{Bounding $\mathbf{R_{3,3}(T)}$} \\

\noindent As in the process of bounding $R_{3,2}(T)$ above, \eqref{eqn:x_rt_rate_of_decay} holds for $\|x_{wt}^{(2)}\|$, replacing $\|x_{rt_k}\|$ with $\|x_{wt_k}\|^{(2)}$ and $\|\hat{\Theta}_{Bt}^{-1}r_t\|$ with $\|\Delta K_tx_{wt}\|$. Thus, with Lemma \ref{lem:gains_bounded}, \eqref{eqn:x_rt_rate_of_decay}, \eqref{eqn:Delta_K_bound}-\eqref{eqn:Delta_Theta_B_bound}, and Lemma \ref{lem:subG_tail_bound} with appropriate union bounds to bound $\max_{t \in [T_{\rm learn}, T)} \|x_{wt_k}^{(2)}\|$ and $\max_{t \in [T_{\rm learn}, T)} \|x_{wt}\|$ we can bound $R_{3,3}(T)$ as
\begin{align}
    R_{3,3}(T) &\leq \sum_{t=T_{\rm learn}}^{T-1} O\left(\|x_{wt}^{(2)}\|\left(\|x_{rt}\| + \|r_t\|\right)\right) \nonumber \\
    &\leq \sum_{k=k_{\rm learn}}^{k_T}\sum_{t=t_k}^{t_{k+1}-1} O\left(\left(d_3^{t-t_k}\|x_{wt_k}^{(2)}\| + \max_{t \in [t_k, t_{k+1})} \|\Delta K_tx_{wt}\|\right)\left(\|x_{rt}\| + \|r_t\|\right)\right) \nonumber \\
    &\leq \sum_{k=k_{\rm learn}}^{k_T} O\left(\|x_{wt_k}^{(2)}\|\right) + \sum_{k=k_{\rm learn}}^{k_T} O\left(T_k\max_{t \in [t_k, t_{k+1})} \|\Delta K_tx_{wt}\|\left(\|x_{rt}\| + \|r_t\|\right)\right) \nonumber \\
    &\leq O\left(\mathrm{polylog}(1 + T)\sum_{k=k_{\rm learn}}^{k_T} \left(1 + T_k(k+1)^{-1}\right)\right). \label{eqn:R_3_3_bound}
\end{align}

\subsubsection*{Putting everything together}

Combining \eqref{eqn:regret_decomposition} with \eqref{eqn:R_1_bound}, \eqref{eqn:R_2_bound}, \eqref{eqn:R_3_decomposition}, \eqref{eqn:R_3_1_bound}, \eqref{eqn:R_3_2_bound}, and \eqref{eqn:R_3_3_bound}, we finally obtain
\begin{equation} \label{eqn:regret_bounds_combined}
    \begin{aligned}
        \Regret(T) &\leq O(1) + \tilde{O}(\sqrt{T}) \\
        &\indenti{=} + \tilde{O}\left(\sum_{k=k_{\rm learn}}^{k_T} \left(1 + (k+1)^{-1/2} + \sqrt{\tilde{O}\left(1 + T_k(k+1)^{-1}\right)} + T_k(k + 1)^{-1}\right)\right).
    \end{aligned}
\end{equation}
The claim follows straightforwardly from \eqref{eqn:regret_bounds_combined} and Lemma \ref{lem:epochs_linear}.
\qed

%% file: Appendices/Additional_Simulations.tex
\section{} \label{app:additional_simulations}

This appendix provides additional simulation details and results beyond those provided in Section \ref{sec:simulations}.

\subsection{Dynamical systems used in the simulations} \label{app:sim_dynamics}

\subsubsection*{Marginally unstable Laplacian system}

The marginally unstable Laplacian dynamics as used by \cite{Dean_2018} are given by
\begin{equation}
    A_* = \begin{bmatrix} 1.01 & 0.01 & 0 \\ 0.01 & 1.01 & 0.01 \\ 0 & 0.01 & 1.01 \end{bmatrix}, \quad B_* = I_3.
\end{equation}
For simulations with this system, $B_*$ is completely known and the uncertainties in $A_*$ are structured as $A_* = I_3 + \Theta_{A*}$. For the case with an initial stabilizing feedback gain, we chose $\hat{B}_0 = B_*$ and $\hat{A}_0 = I_3 + (1 - \Delta)\Theta_{A*}$ for a randomly-selected perturbation $\Delta$, restricted to be sufficiently small in magnitude that the feedback gain $\hat{K}_0 = \mathrm{dlqr}(\hat{A}_0, \hat{B}_0, Q, R)$ stabilized the true plant $(A_*, B_*)$. For the case with an initially unstable feedback gain, we simply chose $\hat{A}_0 = \hat{B}_0 = I_3$. In both cases, the initial reference model was given by $A_m = \hat{A}_0 + \hat{B}_0\hat{K}_0$, $B_m = \hat{B}_0 = B_*$. All simulations with this system have an exogenous noise standard deviation of 0.1.

\subsubsection*{Linearized 6DOF quadrotor}

The quadrotor simulations use the linearized 6-DOF quadrotor model in \citep{Annaswamy2023ACRL} given by
\begin{gather}
    \dot{x} = v_x,\; \dot{v}_x = g\theta,\; \dot{\theta} = q,\; \dot{q} = \frac{1}{I_y}\tau_y, \label{eqn:quadrotor_x_motion} \\
    \dot{y} = v_y,\; \dot{v}_y = -g\phi,\; \dot{\phi} = p,\; \dot{p} = \frac{1}{I_x}\tau_x, \label{eqn:quadrotor_y_motion} \\
    \dot{z} = v_z,\; \dot{v}_z = \frac{1}{m}F, \label{eqn:quadrotor_z_motion} \\
    \dot{\psi} = r,\; \dot{r} = \frac{1}{I_z}\tau_z \label{eqn:quadrotor_psi_motion}
\end{gather}
where $(x, y, z)$ is the COM position, $(v_x, v_y, v_z)$ is the COM velocity, $(\phi, \theta, \psi)$ are roll, pitch, and yaw, $(p, q, r)$ is the angular velocity, $F$ is the net vertical force on the drone, and $(\tau_x, \tau_y, \tau_z)$ are the net torques about each axis. $F$, $\tau_x$, $\tau_y$, and $\tau_z$ are nominally related to the thrusts from each rotor, $u_i$, as
\begin{equation} \label{eqn:quadrotor_forces_thrusts}
    \begin{bmatrix} F \\ \tau_y \\ \tau_x \\ \tau_z \end{bmatrix} = \begin{bmatrix} 1 & 1 & 1 & 1 \\ L & 0 & -L & 0 \\ 0 & L & 0 & -L \\ \nu & -\nu & \nu & -\nu \end{bmatrix}\left(\begin{bmatrix} u_1 \\ u_2 \\ u_3 \\ u_4 \end{bmatrix} - \frac{mg}{4}\begin{bmatrix} 1 \\ 1 \\ 1 \\ 1 \end{bmatrix}\right) = B_{c2}(\vec{u} - \vec{b}_g).
\end{equation}
The quadrotor simulations in this work consider uncertainty in the form of an unknown partial loss of effectiveness on each rotor. Defining $\epsilon_* \in \bbR^4$ as the unknown LOE vector, the actual relationship between $F, \tau_x, \tau_y, \tau_z$, and $u_i$ is given by
\begin{equation}
    \begin{bmatrix} F \\ \tau_y \\ \tau_x \\ \tau_z \end{bmatrix} = B_{c2}(\mathrm{diag}(\epsilon_*)\vec{u} - \vec{b}_g).
\end{equation}
Let $\vec{x} = [x, y, z, \theta, \phi, \psi, v_x, v_y, v_z, q, p, r]^\top$. Then,
\begin{equation} \label{eqn:quadrotor_continuous_dynamics}
    \dot{\vec{x}} = A_c\vec{x} + B_{c1}B_{c2}(\mathrm{diag}(\epsilon_*)\vec{u} - \vec{b}_g)
\end{equation}
where $A_c$ and $B_{c1}$ can be derived from \eqref{eqn:quadrotor_x_motion}-\eqref{eqn:quadrotor_psi_motion}. The physical parameters used were $g = 9.81\mathrm{m/s^2}$, $m = 0.4\mathrm{kg}$, $L = 11.43\mathrm{cm}$, $I_x = I_y = 2.09 * 10^{-3}\mathrm{kg.m^2}$, $I_z = 4.18 * 10^{-3}\mathrm{kg.m^2}$, and $\nu = 1.524\mathrm{cm}$, and the actual LOE used in the simulations was $\epsilon_* = [0.5, 1, 1, 1]^\top$, which would result in the quadrotor both losing altitude and pitching downward if not corrected by an adaptive controller.

As a final step, the dynamics in \eqref{eqn:quadrotor_continuous_dynamics} were Euler discretized with a time step of $\Delta t = 0.01$ seconds to yield the discrete-time dynamics
\begin{equation} \label{eqn:quadrotor_discrete_dynamics}
    \begin{gathered}
        \vec{x}_{t+1} = A\vec{x}_t + B_m(\Theta_{B*}\vec{u}_t - \vec{b}_g) + \vec{w}_{t+1}, \\
        A = I_{12} + \Delta tA_c, \quad B_m = \Delta tB_{1c}B_{2c}, \quad \Theta_{B*} = \mathrm{diag}(\epsilon_*).
    \end{gathered}
\end{equation}
The initial gain was given by $\hat{K}_0 = \mathrm{dlqr}(A, B_m, Q, R)$, the initial reference dynamics were given by $A_m = A + B_m\hat{K}_0$, and at every time step, an extra term $\hat{\Theta}_{Bt}^{-1}\vec{b}_g$ was added to $\vec{u}_t$. Simulations with this system have an exogenous noise standard deviation of 0.01.


\subsection{Additional simulations using Gaussian exploration} \label{app:gaussian_sims}

\begin{figure}[h]
    \centering
    \begin{subfigure}[t]{0.49\textwidth}
        \centering
        \caption{Regret}
        \label{fig:laplacian_stable_explore_0.01_regret}
        \includegraphics[width=\textwidth]{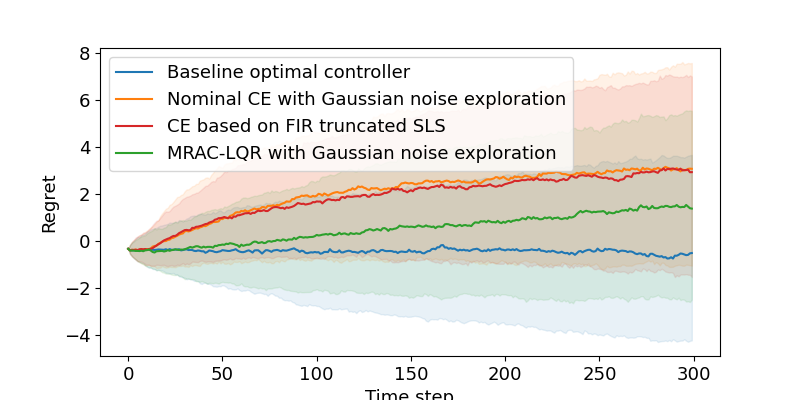}
    \end{subfigure}
    \hfill
    \begin{subfigure}[t]{0.49\textwidth}
        \centering
        \caption{State magnitude}
        \label{fig:laplacian_stable_explore_0.01_state}
        \includegraphics[width=\textwidth]{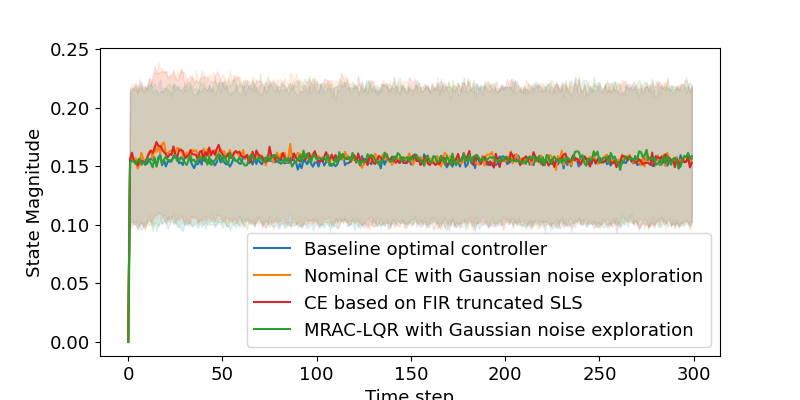}
    \end{subfigure}
    \caption{Laplacian system with stable initial controller and $\sigma_{\rm explore} = 0.01$. Solid lines are the median values over 1000 trials, and shaded regions are the 20\%-80\% confidence windows.}
    \label{fig:laplacian_stable_explore_0.01}
\end{figure}

Figure \ref{fig:laplacian_stable_explore_0.01} shows simulation results for the Laplacian system with an initial stabilizing controller but very little exploration. Even when an initial stabilizing controller is given, indirect adaptive control can still perform poorly cost-wise if there is not enough excitation to learn effectively, and we see in Figure \ref{fig:laplacian_stable_explore_0.01} that MRAC-LQR outperforms other methods in this scenario due to its active stabilization at every time step.

\begin{figure}[h]
    \centering
    \begin{subfigure}[t]{0.49\textwidth}
        \centering
        \caption{Regret}
        \label{fig:quadrotor_low_noise_regret}
        \includegraphics[width=\textwidth]{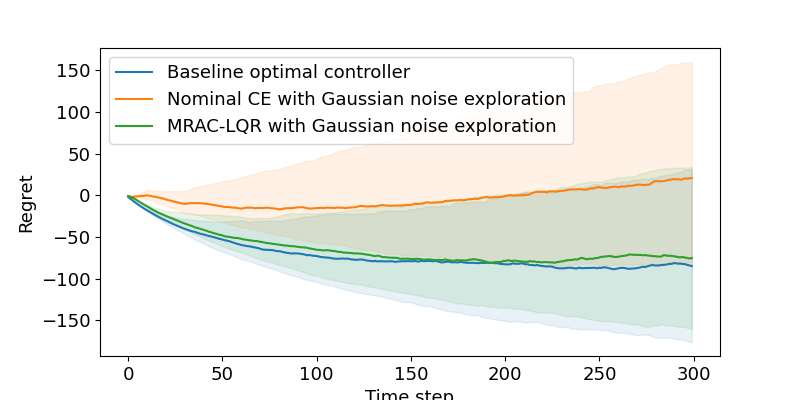}
    \end{subfigure}
    \hfill
    \begin{subfigure}[t]{0.49\textwidth}
        \centering
        \caption{Pitch angle magnitude}
        \label{fig:quadrotor_low_noise_state}
        \includegraphics[width=\textwidth]{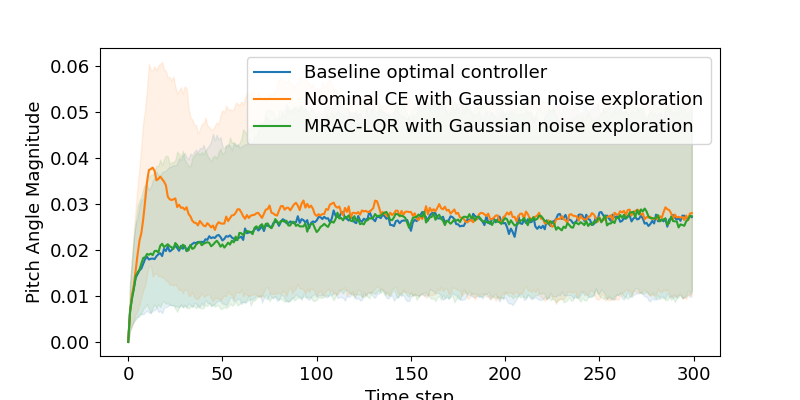}
    \end{subfigure}
    \caption{6DOF quadrotor with $\sigma_{\rm explore} = 0.01$. Solid lines are the median values over 1000 trials, and shaded regions are the 20\%-80\% confidence windows.}
    \label{fig:quadrotor_low_noise}
\end{figure}

Figure \ref{fig:quadrotor_low_noise} shows simulation results for the quadrotor, which will fall out of the air in the absence of adaptation and is thus open-loop unstable with an unstable initial controller. Here, we compare MRAC-LQR only to the baseline optimal controller and to the nominal CE approach, which was consistently the best-performing competitor to MRAC-LQR in our simulations. We see that MRAC-LQR is able to stabilize faster than the nominal CE approach, keeping the quadrotor more level and the accumulated cost lower.

\subsection{Additional simulations using deterministic exploration} \label{app:PE_sims}

\begin{figure}[h]
    \centering
    \begin{subfigure}[t]{0.49\textwidth}
        \centering
        \caption{Regret}
        \label{fig:PE_laplacian_stable_explore_0.1_regret}
        \includegraphics[width=\textwidth]{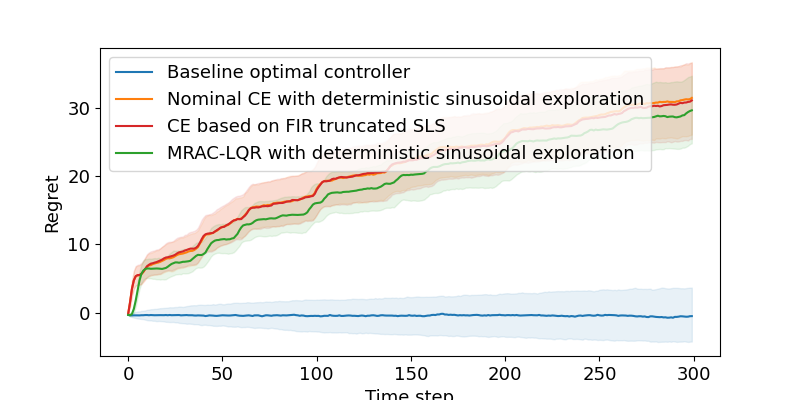}
    \end{subfigure}
    \hfill
    \begin{subfigure}[t]{0.49\textwidth}
        \centering
        \caption{State magnitude}
        \label{fig:PE_laplacian_stable_explore_0.1_state}
        \includegraphics[width=\textwidth]{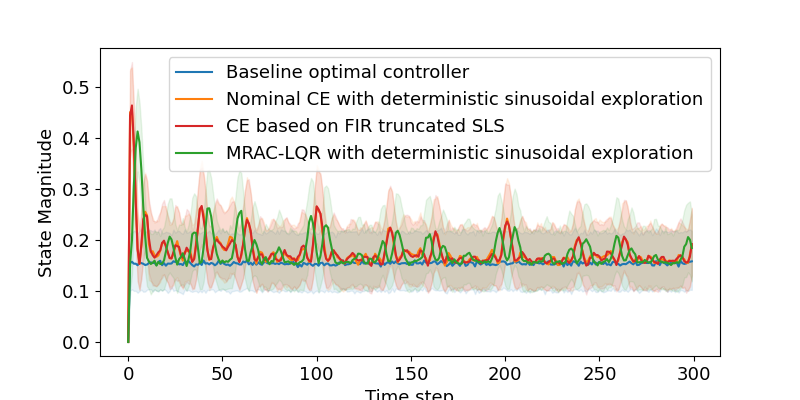}
    \end{subfigure}
    \caption{Laplacian system with stable initial controller and deterministic sinusoidal exploration with $\sigma_{\rm explore} = 0.1$. Solid lines are the median values over 1000 trials, and shaded regions are the 20\%-80\% confidence windows.}
    \label{fig:PE_laplacian_stable_explore_0.1}
\end{figure}

\begin{figure}[h]
    \centering
    \begin{subfigure}[t]{0.49\textwidth}
        \centering
        \caption{Regret}
        \label{fig:PE_laplacian_unstable_explore_0.1_regret}
        \includegraphics[width=\textwidth]{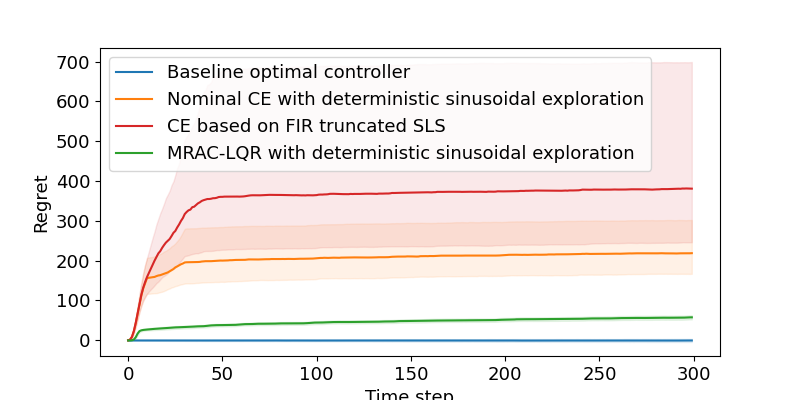}
    \end{subfigure}
    \hfill
    \begin{subfigure}[t]{0.49\textwidth}
        \centering
        \caption{State magnitude}
        \label{fig:PE_laplacian_unstable_explore_0.1_state}
        \includegraphics[width=\textwidth]{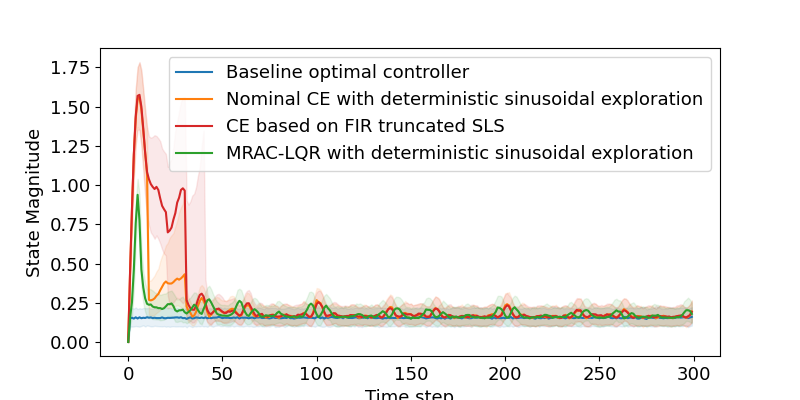}
    \end{subfigure}
    \caption{Laplacian system with unstable initial controller and deterministic sinusoidal exploration with $\sigma_{\rm explore} = 0.1$. Solid lines are the median values over 1000 trials, and shaded regions are the 20\%-80\% confidence windows.}
    \label{fig:PE_laplacian_unstable_explore_0.1}
\end{figure}

In this section, we repeat the simulations in Figures \ref{fig:laplacian_stable_explore_0.1} and \ref{fig:laplacian_unstable_explore_0.1} using a deterministic sinusoidal exploration as in Line \ref{algln:r_t} of Algorithm \ref{alg:mrac-lqr}, rather than Gaussian noise exploration as in the previous simulations. Comparing Figures \ref{fig:laplacian_stable_explore_0.1} and \ref{fig:PE_laplacian_stable_explore_0.1}, both of which use an initial stabilizing controller, we see that the sinusoidal exploration tends to learn slightly more slowly and thus accumulate slightly more regret than Gaussian noise exploration in this pristine setting with no unmodeled dynamics. However, Figures \ref{fig:PE_laplacian_stable_explore_0.1} and \ref{fig:PE_laplacian_unstable_explore_0.1} show that the ranking of the algorithms' performance is preserved: if all algorithms employ deterministic sinusoidal exploration, MRAC-LQR retains its advantages over the other approaches. In fact, comparing Figures \ref{fig:laplacian_unstable_explore_0.1} and \ref{fig:PE_laplacian_unstable_explore_0.1}, we see that the other approaches suffer a large cost increase from lack of an initial stabilizing controller and a slower method of exploration, while MRAC-LQR is much less impacted.

It is worth bearing in mind that, as shown by \cite{sarker2023accurate}, there is an important practical reason to apply deterministic sinusoidal exploration rather than using Gaussian noise. Although the simulations in this paper do not include unmodeled dynamics, they are always present in practice, and typically become prevalent only at high frequencies. When applying deterministic sinusoids, one can choose the frequency content to avoid exciting the unmodeled dynamics, which cannot be done with the Gaussian approach.

%% file: references.bib
@Book{Goodwin_1984,
  title     = {Adaptive Filtering Prediction and Control},
  publisher = {Prentice Hall},
  year      = {1984},
  author    = {Graham C Goodwin and Kwai Sang Sin},
}

@InCollection{Dean_2018,
  author    = {Sarah Dean and Horia Mania and Nikolai Matni and Benjamin Recht and Stephen Tu},
  title     = {Regret Bounds for Robust Adaptive Control of the Linear Quadratic Regulator},
  booktitle = {Advances in Neural Information Processing Systems 31},
  publisher = {Curran Associates, Inc.},
  year      = {2018},
  pages     = {4192--4201},
}

@Book{Landau11,
  title={Adaptive Control: Algorithms, Analysis and Applications},
  author={Landau, Ioan Dor{\'e} and Lozano, Rogelio and M'Saad, Mohammed and Karimi, Alireza},
  year={2011},
  publisher={Springer Science \& Business Media}
}

@Article{Campi98,
  author    = {Campi, M C. and Kumar, P R.},
  journal   = {SIAM Journal on Control and Optimization},
  title     = {Adaptive linear quadratic gaussian control: the cost-biased approach revisited},
  year      = {1998},
  number    = {6},
  pages     = {1890--1907},
  volume    = {36},
  publisher = {SIAM},
}

@Book{Narendra05,
  author    = {Kumpati S. Narendra and Anuradha M. Annaswamy},
  publisher = {Dover Publications},
  title     = {Stable Adaptive Systems},
  year      = {2005},
  address   = {NJ},
  note      = {(original publication by Prentice-Hall Inc., 1989)},
}

@inproceedings{abbasi2011,
  title={Regret bounds for the adaptive control of linear quadratic systems},
  author={Abbasi-Yadkori, Yasin and Szepesv{\'a}ri, Csaba},
  booktitle={Proceedings of the 24th Annual Conference on Learning Theory},
  pages={1--26},
  year={2011},
  organization={JMLR Workshop and Conference Proceedings}
}

@article{sarker2023accurate,
  title={Accurate parameter estimation for safety-critical systems with unmodeled dynamics},
  author={Arnab Sarker and Peter Fisher and Joseph E. Gaudio and Anuradha M. Annaswamy},
  journal={Artificial Intelligence},
  pages={103857},
  year={2023},
  publisher={Elsevier}
}

@article{annaswamy2023arcra,
  title={Adaptive Control and Intersections with Reinforcement Learning},
  author={Annaswamy, Anuradha M},
  journal={Annual Review of Control, Robotics, and Autonomous Systems},
  volume={6},
  year={2023},
  publisher={Annual Reviews}
}

@inproceedings{lale2022reinforcement,
  title={Reinforcement learning with fast stabilization in linear dynamical systems},
  author={Lale, Sahin and Azizzadenesheli, Kamyar and Hassibi, Babak and Anandkumar, Animashree},
  booktitle={International Conference on Artificial Intelligence and Statistics},
  pages={5354--5390},
  year={2022},
  organization={PMLR}
}

@INPROCEEDINGS{Miller2017,
  author={Miller, Daniel E.},
  booktitle={2017 IEEE Conference on Control Technology and Applications (CCTA)}, 
  title={Classical discrete-time adaptive control revisited: Exponential stabilization}, 
  year={2017},
  volume={},
  number={},
  pages={1975-1980},
  doi={10.1109/CCTA.2017.8062745}
}

@article{Kreisselmeier1986,
title = {Adaptive control of a class of slowly time-varying plants},
journal = {Systems \& Control Letters},
volume = {8},
number = {2},
pages = {97-103},
year = {1986},
issn = {0167-6911},
doi = {https://doi.org/10.1016/0167-6911(86)90067-8},
url = {https://www.sciencedirect.com/science/article/pii/0167691186900678},
author = {Gerhard Kreisselmeier}
}

@article{Annaswamy2023ACRL,
  author={Annaswamy, Anuradha M. and Guha, Anubhav and Cui, Yingnan and Tang, Sunbochen and Fisher, Peter A. and Gaudio, Joseph E.},
  journal={IEEE Transactions on Automatic Control}, 
  title={Integration of Adaptive Control and Reinforcement Learning for Real-time Control and Learning}, 
  year={2023},
  volume={},
  number={},
  pages={1-16},
  doi={10.1109/TAC.2023.3290037}
}

@article{astrom1973selftuning,
  title = {On self tuning regulators},
  journal = {Automatica},
  volume = {9},
  number = {2},
  pages = {185-199},
  year = {1973},
  issn = {0005-1098},
  doi = {https://doi.org/10.1016/0005-1098(73)90073-3},
  url = {https://www.sciencedirect.com/science/article/pii/0005109873900733},
  author = {K.J. Åström and B. Wittenmark}}

@article{chen1986leastsquares_lowercase,
  author = {Han-Fu Chen and Lei Guo},
  title = {Convergence rate of least-squares identification and adaptive control for stochastic systems},
  journal = {International Journal of Control},
  volume = {44},
  number = {5},
  pages = {1459--1476},
  year = {1986},
  publisher = {Taylor \& Francis},
  doi = {10.1080/00207178608933679},
  URL = {https://doi.org/10.1080/00207178608933679},
  eprint = {https://doi.org/10.1080/00207178608933679}}

@inproceedings{mania2019CEefficient,
  author = {Mania, Horia and Tu, Stephen and Recht, Benjamin},
  booktitle = {Advances in Neural Information Processing Systems},
  editor = {H. Wallach and H. Larochelle and A. Beygelzimer and F. d\textquotesingle Alch\'{e}-Buc and E. Fox and R. Garnett},
  pages = {},
  publisher = {Curran Associates, Inc.},
  title = {Certainty Equivalence is Efficient for Linear Quadratic Control},
  url = {https://proceedings.neurips.cc/paper\_files/paper/2019/file/5dbc8390f17e019d300d5a162c3ce3bc-Paper.pdf},
  volume = {32},
  year = {2019}}

@InProceedings{simchowitz2020naive,
  title =    {Naive Exploration is Optimal for Online {LQR}},
  author =       {Simchowitz, Max and Foster, Dylan},
  booktitle =    {Proceedings of the 37th International Conference on Machine Learning},
  pages =    {8937--8948},
  year =   {2020},
  editor =   {III, Hal Daumé and Singh, Aarti},
  volume =   {119},
  series =   {Proceedings of Machine Learning Research},
  month =    {13--18 Jul},
  publisher =    {PMLR},
  url =    {https://proceedings.mlr.press/v119/simchowitz20a.html}
}

@inproceedings{ibrahimi2012sparseadaptiveLQR,
  author = {Ibrahimi, Morteza and Javanmard, Adel and Roy, Benjamin},
  booktitle = {Advances in Neural Information Processing Systems},
  editor = {F. Pereira and C.J. Burges and L. Bottou and K.Q. Weinberger},
  pages = {},
  publisher = {Curran Associates, Inc.},
  title = {Efficient Reinforcement Learning for High Dimensional Linear Quadratic Systems},
  url = {https://proceedings.neurips.cc/paper\_files/paper/2012/file/a9eb812238f753132652ae09963a05e9-Paper.pdf},
  volume = {25},
  year = {2012}
}

@InProceedings{cohen2019SDPdare,
  title =    {Learning Linear-Quadratic Regulators Efficiently with only $\sqrt{T}$ Regret},
  author =       {Cohen, Alon and Koren, Tomer and Mansour, Yishay},
  booktitle =    {Proceedings of the 36th International Conference on Machine Learning},
  pages =    {1300--1309},
  year =   {2019},
  editor =   {Chaudhuri, Kamalika and Salakhutdinov, Ruslan},
  volume =   {97},
  series =   {Proceedings of Machine Learning Research},
  month =    {09--15 Jun},
  publisher =    {PMLR},
  url =    {https://proceedings.mlr.press/v97/cohen19b.html}}

@INPROCEEDINGS{bradtke1994adaptiveLQR,
  author={Bradtke, S.J. and Ydstie, B.E. and Barto, A.G.},
  booktitle={Proceedings of 1994 American Control Conference - ACC '94}, 
  title={Adaptive linear quadratic control using policy iteration}, 
  year={1994},
  volume={3},
  number={},
  pages={3475-3479 vol.3},
  doi={10.1109/ACC.1994.735224}}

@article{jiang2012adaptiveoptimal,
  title = {Computational adaptive optimal control for continuous-time linear systems with completely unknown dynamics},
  journal = {Automatica},
  volume = {48},
  number = {10},
  pages = {2699-2704},
  year = {2012},
  issn = {0005-1098},
  doi = {https://doi.org/10.1016/j.automatica.2012.06.096},
  url = {https://www.sciencedirect.com/science/article/pii/S0005109812003664},
  author = {Yu Jiang and Zhong-Ping Jiang}}

@InProceedings{fazel2018policygradient,
  title =    {Global Convergence of Policy Gradient Methods for the Linear Quadratic Regulator},
  author =       {Fazel, Maryam and Ge, Rong and Kakade, Sham and Mesbahi, Mehran},
  booktitle =    {Proceedings of the 35th International Conference on Machine Learning},
  pages =    {1467--1476},
  year =   {2018},
  editor =   {Dy, Jennifer and Krause, Andreas},
  volume =   {80},
  series =   {Proceedings of Machine Learning Research},
  month =    {10--15 Jul},
  publisher =    {PMLR},
  url =    {https://proceedings.mlr.press/v80/fazel18a.html}}

@ARTICLE{mohammadi2021dlqr,
  author={Mohammadi, Hesameddin and Soltanolkotabi, Mahdi and Jovanović, Mihailo R.},
  journal={IEEE Control Systems Letters}, 
  title={On the Linear Convergence of Random Search for Discrete-Time LQR}, 
  year={2021},
  volume={5},
  number={3},
  pages={989-994},
  doi={10.1109/LCSYS.2020.3006256}}

@ARTICLE{hjalmarsson1998ift,
  author={Hjalmarsson, H. and Gevers, M. and Gunnarsson, S. and Lequin, O.},
  journal={IEEE Control Systems Magazine}, 
  title={Iterative feedback tuning: theory and applications}, 
  year={1998},
  volume={18},
  number={4},
  pages={26-41},
  doi={10.1109/37.710876}}

@INPROCEEDINGS{pang2018datadriven,
  author={Pang, Bo and Bian, Tao and Jiang, Zhong-Ping},
  booktitle={2018 IEEE Conference on Decision and Control (CDC)}, 
  title={Data-driven Finite-horizon Optimal Control for Linear Time-varying Discrete-time Systems}, 
  year={2018},
  volume={},
  number={},
  pages={861-866},
  doi={10.1109/CDC.2018.8619347}}

@INPROCEEDINGS{berberich2020datadriven,
  author={Berberich, Julian and Koch, Anne and Scherer, Carsten W. and Allg{\"o}wer, Frank},
  booktitle={2020 American Control Conference (ACC)}, 
  title={Robust data-driven state-feedback design}, 
  year={2020},
  volume={},
  number={},
  pages={1532-1538},
  doi={10.23919/ACC45564.2020.9147320}}

@misc{vanwaarde2020datainformativity,
  title={Data informativity: a new perspective on data-driven analysis and control}, 
  author={Henk J. van Waarde and Jaap Eising and Harry L. Trentelman and M. Kanat Camlibel},
  year={2020},
  eprint={1908.00468},
  archivePrefix={arXiv},
  primaryClass={math.OC},
  url={https://arxiv.org/abs/1908.00468}}

@ARTICLE{dorfler2023datadriven,
  author={D{\"o}rfler, Florian and Tesi, Pietro and De Persis, Claudio},
  journal={IEEE Transactions on Automatic Control}, 
  title={On the Certainty-Equivalence Approach to Direct Data-Driven LQR Design}, 
  year={2023},
  volume={68},
  number={12},
  pages={7989-7996},
  doi={10.1109/TAC.2023.3253787}
  }

@ARTICLE{guo1996WLSAdaptiveControl,
  author={Lei Guo},
  journal={IEEE Transactions on Automatic Control}, 
  title={Self-convergence of weighted least-squares with applications to stochastic adaptive control}, 
  year={1996},
  volume={41},
  number={1},
  pages={79-89},
  doi={10.1109/9.481609}
}

@book{Vershynin2019,
  title={High-dimensional probability: An introduction with applications in data science},
  author={Vershynin, Roman},
  volume={47},
  year={2018},
  publisher={Cambridge university press}
}

@Article{Pisier2016,
  author    = {Gilles Pisier},
  title     = {Subgaussian sequences in probability and Fourier analysis},
  journal   = {Graduate J. Math},
  year      = {2016},
  volume    = {1},
  pages     = {60--80}
}

@book{stengel1994optimal,
  title={Optimal control and estimation},
  author={Stengel, Robert F},
  year={1994},
  publisher={Courier Corporation}
}

@article{lee1998DAREBounds,
	title={Solution Bounds for the Discrete Riccati Equation and Its Applications},
	author={C. H. Lee and Y. C. Chang},
	year={1998},
	journal={Journal of Optimization Theory and Applications},
	volume={99},
	number={},
	pages={443-463},
	doi={10.1023/A:1021730512293},
}

@book{hespanha2018LinearSystemsTheory,
 ISBN = {9780691179575},
 author = {João P. Hespanha},
 edition = {NED - New edition, 2},
 publisher = {Princeton University Press},
 title = {Linear Systems Theory: Second Edition},
 urldate = {2026-02-14},
 year = {2018}
}

@mastersthesis{fisher2023MastersThesis,
author={Fisher,Peter},
year={2023},
title={Fast Adaptive Laws for Adaptive Control Under Stochastic Disturbances},
journal={ProQuest Dissertations and Theses},
pages={172},
isbn={9798381957334},
language={English},
url={https://www.proquest.com/dissertations-theses/fast-adaptive-laws-control-under-stochastic/docview/3031024369/se-2},
school={Massachusetts Institute of Technology}
}

@book{chen1991Identification,
  title={Identification and Stochastic Adaptive Control},
  author={Han-Fu Chen and Lei Guo},
  year={1991},
  publisher={Birkh{\"a}user Boston},
  doi={https://doi.org/10.1007/978-1-4612-0429-9}
}

@article{ayoub1974EulerZeta,
author = {Raymond Ayoub},
title = {Euler and the Zeta Function},
journal = {The American Mathematical Monthly},
volume = {81},
number = {10},
pages = {1067--1086},
year = {1974},
publisher = {Taylor \& Francis},
doi = {10.1080/00029890.1974.11993738},
URL = {https://doi.org/10.1080/00029890.1974.11993738},
eprint = {https://doi.org/10.1080/00029890.1974.11993738}
}
